\documentclass{aa}  
\usepackage{natbib}
\bibpunct{(}{)}{;}{a}{}{,} 
\usepackage{graphicx}
\usepackage{txfonts}

\usepackage[dvipsnames]{xcolor}
\usepackage{amsmath}
\usepackage{hyperref}
\hypersetup{
    colorlinks=true,
    citecolor=blue,
    linkcolor=blue,
    filecolor=magenta,      
    urlcolor=cyan}
\usepackage{caption}
\usepackage{lipsum}

\begin{document}

   \title{Cosmography by orthogonalized  logarithmic polynomials}
   \titlerunning{Logarithmic Cosmography}
   \authorrunning{Bargiacchi G. et al.}


   \author{G. Bargiacchi 
          \inst{1}\fnmsep\thanks{e-mail: giada.bargiacchi@unina.it},
          G. Risaliti\inst{2,3},
          M. Benetti\inst{1,4,5},
          S. Capozziello\inst{1,4,5,6},
          E. Lusso\inst{2,3},
          A. Saccardi\inst{2},
          M. Signorini\inst{2,3}}
\institute{Scuola Superiore Meridionale, Largo S. Marcellino 10, I-80138, Napoli,\and
   Dipartimento di Fisica e Astronomia, Università degli Studi di Firenze, via G. Sansone 1, 50019 Sesto Fiorentino, Firenze, Italy \and Istituto Nazionale di Astrofisica (INAF) – Osservatorio Astrofisico di Arcetri, 50125 Florence, Italy, \and
Dipartimento di Fisica "E. Pancini", Università degli Studi di Napoli "Federico II, Complesso Univ. Monte S. Angelo, Via Cinthia 9, I-80126, Napoli,\and 
Istituto Nazionale di Fisica Nucleare (INFN), Sez. di Napoli, Complesso Univ. Monte S. Angelo, Via Cinthia 9, I-80126, Napoli,\and 
Laboratory for Theoretical Cosmology, Tomsk State University of Control Systems and Radioelectronics (TUSUR), 634050 Tomsk,  Russia.}

   \date{Received January 15, 2021; accepted XX}

  \abstract
{Cosmography is a powerful tool to investigate the Universe kinematic and then to reconstruct  dynamics in a model-independent way. However, recent new measurements of supernovae Ia and quasars have populated the Hubble diagram up to high redshifts ($z \sim 7.5$) and the application of the traditional cosmographic approach has become less straightforward due to the large redshifts implied. Here we investigate this issue through an expansion of the luminosity distance-redshift relation in terms of "orthogonal" logarithmic polynomials. In particular we point out the advantages of a new procedure of "orthogonalization" and we show that such an expansion provides a very good fit in the whole $z=0\div 7.5$ range to both real and mock data obtained assuming various cosmological models. Moreover, despite of the fact that the cosmographic series is tested well beyond its convergence radius, the parameters obtained expanding the luminosity distance - redshift relation for the $\Lambda$CDM model are broadly consistent with the results from a fit of mock data obtained with the same cosmological model. This provides a method to test the reliability of a cosmographic function to study cosmological models at high redshifts and it demonstrates that the logarithmic polynomial series can be used to test the consistency of the $\Lambda$CDM model with the current Hubble diagram of quasars and supernovae Ia. We confirm a strong tension (at $>4\sigma$) between the concordance cosmological model and the Hubble diagram at $z>1.5$. Such a tension is dominated by the  contribution of quasars at $z>2$ and starts to be present also in the few supernovae Ia observed at $z>1$.}


\maketitle
%
\section{Introduction}

The Universe is in a phase of accelerated expansion, as observed by cosmological probes such as local type Ia supernovae (SNe Ia, see e.g. \citealt{riess1998,perlmutter1999}) and the combination of SNeIa, Baryonic Acoustic Oscillations (BAO), Cosmic Microwave Background (CMB), and lensing provided by \citet{planck2018}. This acceleration is generally ascribed to the so-called \textit{dark energy}. According to the concordance cold dark matter model ($\mathrm{\Lambda}$CDM), this dark energy component is naively associated to  a cosmological constant ($\mathrm{\Lambda}$) generated, in principle,  by the quantum fluctuations of vacuum \citep{2000IJMPD...9..373S,2001LRR.....4....1C,2003RvMP...75..559P} and represents the $\sim 70\%$ of the total energy, being dominant in the current Universe. However the huge discrepancy between vacuum fluctuations of primordial universe and the today observed value of $\Lambda$ poses dramatically the so called {\it Cosmological Constant Problem} (\citealt{Weinberg}).
The flat $\mathrm{\Lambda}$CDM model is the one that, based on the simplest theory, can reproduce a large number of observations. Nevertheless, it suffers from both theoretical (e.g. \textit{fine-tuning} and \textit{coincidence} problems) and observational issues (e.g. the missing satellites problems, \citealt{2010arXiv1009.4505B}). As a further example of the latter, we can mention the recent tension on the Hubble constant $H_{0}$, which is the local value of the Hubble parameter $H(z)$. Indeed, the value of $H_{0}$ inferred from the CMB, through the assumption of the flat $\mathrm{\Lambda}$CDM model, is in tension at more than $4\sigma$ with the one measured from local sources (Cepheids, SNe Ia), as reported in \citet{riess2019}, and with the one obtained in \citet{2020MNRAS.tmp.1661W} from lensed quasars. If we rule out systematic errors in these measurements, new physics beyond the standard cosmological model is needed (e.g. time-dependent dark energy equation
of state, modified gravity, additional relativistic particles and so on; see \citealt{RoccoReview,Spallicci,2020ApJ...894...54D,2020arXiv201213932A,2020PhRvD.101l3516A,2020arXiv201110559R}).\\
Model-independent approaches have been proposed to investigate the correct form of dark energy without any assumption on its physical origin or the constituents of the Universe. Among these techniques, \textit{cosmography} \citep{1972gcpa.book.....W} is one of the most consolidated. It relies only on the homogeneity and isotropy of the Universe without requiring any explicit physical function for the scale factor $a(t)$, but only an analytical expression for it. The major problem of cosmography is that, to detect possible deviations from the standard cosmological model, data at high redshifts (i.e. $z>2$) are necessary and convergence issues emerge in this redshift range \citep{RoccoHigh}. Indeed, if the data exceed the limit $t \sim t_{0}$ ($z \sim 0$), where $t_{0}$ is the present epoch, it is not correct to apply a Taylor expansion around $t_{0}$, the one traditionally used in the cosmographic approach for the analytical approximation of $a(t)$ \citep{visser2004,RoccoReview}. By definition of the Taylor expansion, this standard cosmographic approach gives an approximation of the luminosity distance $D_{L}(z)$ which is correct only around $t=t_{0}$, in the limit of $z\leq1$. As a consequence, the cosmographic Taylor expansion cannot be used to analyze data at $z\geq 1$. This was not a problem until recently as the only standard candles in the Hubble diagram were SNe  observed up to $z \sim 1$ \citep{betoule2014} for which the Taylor approach was completely justified. However, new SNe  have been observed up to $z=2.26$ by the \textit{Pantheon} survey \citep{scolnic2018} and our research group has developed a method to turn quasars into standard candles \citep{rl19,salvestrini2019,lr16,rl15,2020A&A...642A.150L} extending the Hubble diagram up to $z \sim 7.5$ \citep{banados2018}. Thanks to this extended redshift range, the joint sample of quasars and SNe allows us to effectively look for discrepancies between the data and the standard cosmological model (see e.g. \citealt{2020ApJ...900...70R}), but this analysis could be possible only solving the convergence issues described above. Some techniques have been recently proposed to alleviate this problem. In particular, the use of rational polynomials is capable of improving the convergence radius and then to approach higher redshift ranges as in the case of Pad\'e and Chebyshev polynomials (see e.g. \citealt{Aviles,Chebyshev,2017A&A...598A.113D,2020JCAP...12..007Z}). \\
In this paper, we propose a novel technique.  We try to overcome the weakness of cosmography at high redshifts by substituting the Taylor expansion with a new analytical function. We expand the luminosity distance in orthogonal polynomials of logarithmic functions. We investigate the advantage of this cosmographic technique in reproducing and testing cosmological models, and fitting the Hubble diagram of quasars and SNe. Once proven the robustness of the method, we also apply it to test the flat $\mathrm{\Lambda}$CDM model.\\

The paper is structured as follows. In Section \ref{section2} we introduce the main features of cosmography, our logarithmic expansion and the "orthogonalization" procedure we are going to apply. Section \ref{section3} is focused on the basic demands of cosmography and on how the logarithmic expansion handles with them. This accurate analysis is also motivated by other cosmographic studies on the logarithmic approach (see e.g. \citealt{2020arXiv200904109B,2020PhRvD.102l3532Y}). Indeed, this section aims at demonstrating that, in the redshift range analyzed in this work, our approach can be used to give a correct analytical reproduction of data and cosmological models and to test physical models. We then use this method as a cosmographic test of cosmological models in Section \ref{section4}. In particular, we test the flat $\mathrm{\Lambda}$CDM model through a joint fit\footnote{The fits presented in this work are all obtained by employing a fully Bayesian Monte Carlo Markov Chain (MCMC) based approach. Specifically, we considered the affine-invariance MCMC Python package \textit{emcee} \citep{2013PASP..125..306F}, where the estimated uncertainties on the parameters are taken into account. This technique has been already adopted in \cite{Salzano} .} of the Hubble diagram of quasars and SNe, confirming the tension with the flat $\mathrm{\Lambda}$CDM at more than $4\sigma$. Finally, in Section \ref{section5} we draw conclusions and perspectives of our work. In Appendix \ref{appendixk}, we explain how to properly perform the "ortoghonalization" procedure in the logarithmic function. In Appendix \ref{appendixcoeff} we give the complete formulae for the fourth and fifth-order logarithmic expansions in relation with a flat $\mathrm{\Lambda}$CDM model.\\

Throughout the paper, we use natural units with $c=1$. Moreover, we assume a curvature $k=0$ consistently with the most recent cosmological observations on CMB \citep{planck2018}.

\section{The cosmographic approach}
\label{section2}

The cosmographic approach is based only on the isotropy and homogeneity of the Universe through the assumption of the Friedmann-Lema\^{i}tre-Robertson-Walker (FLRW) metric \citep{1972gcpa.book.....W}. This is a purely geometrical description of the Universe kinematic  in which all the physics is hidden in the scale factor $a(t)$. A cosmographic approach does not depend on any cosmological assumption, besides the Cosmological Principle,  as it does not require any equation of state of the cosmic fluid postulated \textit{a priori} but only an analytical expression for the scale factor $a(t)$ in the FLRW metric. This technique allows to study the evolution of the Universe in a complete model-independent way.\\
In general, such an approach has two main advantages: 1) if the adopted expansion is sufficiently flexible, it is capable to fit the observational data with high accuracy; 2) the cosmographic parameters can be used to test any cosmological model \citep{Chebyshev,2019IJMPD..2850154E}. This is done by expanding the chosen cosmological model and the cosmographic series, which gives the relations between the physical parameters of the model and the cosmographic parameters \citep{Aviles,RoccoReview,Benetti}. Then the values of these parameters are compared with the results of the cosmographic fit. An example of this approach is the fit of the Hubble diagram of SNe with a second-order Taylor expansion, whose free parameters are the Hubble constant $H_{0}$ and the deceleration parameter $q_0$ \citep{2004ApJ...607..665R}. The negative value of $q_0$ obtained in this analysis provides the fundamental information on the present acceleration of the expansion of the Universe. Moreover, when $q_0$ is expressed as a function of $\Omega_{M,0}$ in a flat $\Lambda$CDM model, we obtain $\Omega_{M,0}\sim0.3$, and therefore $\Omega_{\Lambda,0}\sim0.7$. A further step in this approach is the inclusion of a third-order term in the cosmographic expansion, where the additional free parameter is the so-called "jerk" \citep{visser2004} which is expected to be equal to 1 in a flat $\Lambda$CDM model. The fit of this third-order polynomial not only provides a more precise determination of $q_0$, but it is also a test for the model itself: any significant deviation from 1 of the best fit value of the jerk would be an indication of a tension between the data and the flat $\Lambda$CDM model (see e.g. \citealt{2017A&A...598A.113D}). \\
In principle, this method can be used to test any cosmological model. However, as already stated in the previous section, in the past few years its application has become less straightforward, due the increased redshift range of the observational data. 
This has introduced two new problems. First, the number of terms needed to reproduce the observational data with a cosmographic expansion has increased to four or even five \citep{2017A&A...598A.113D,RoccoHigh,Benetti}, which makes the estimate of the significance of each parameter (and the comparison with the expectations of physical models) complicated due to the degeneracy among the cosmographic parameters. Second, even if the degeneracy issues are solved, it is not guaranteed that the results of a cosmographic fit can be compared with the theoretical prediction of a given physical model which is calculated around $z=0$ (see e.g. \citealt{Anjan}).\\

The work we present in this paper is aimed at investigating and solving these two issues, in order to apply the cosmographic method to the complete Hubble diagram of SNe and quasars. In this section, we introduce a cosmographic function consisting of "orthogonal" logarithmic polynomials, which solves the former problem. In Section \ref{section3}, we discuss the second problem and validate our method in the whole z=0-7.5 redshift range. 

\subsection{Cosmographic logarithmic approach and "orthogonalization" procedure}
\label{sectionlog}

The cosmographic technique we propose is an extension of the one already described in some  previous works \citep{rl19,lusso19}. It is based on an expansion of the luminosity distance as a power series of the quantity $\log _{10}(1+z)$. With respect to our previous works, we modified the expression of $D_{L}(z)$ to make the cosmographic coefficients "orthogonal"\footnote{Through this paper with the term "orthogonal" we mean  "with no covariance".}. This makes our method unique: in the other cosmographic approaches all the coefficients are correlated as they are combinations of the ones at lower orders. The main advantage of the orthogonal polynomials is that a change in the truncation order of the series does not change the values of the cosmographic coefficients $a_{i}$ as they are uncorrelated. This allows us to test the significance of a possible additional term in the expansion (through its deviation from zero) and to evaluate the tension of the cosmographic fit with a given point in the parameter space without taking into account the correlation amongst free parameters.\\
We estimated the fifth-order for the orthogonal logarithmic polynomial expansion as follows\footnote{For sake of simplicity we use $\text{log}$ instead of $\text{log}_{10}$ and $\text{ln}$ instead of $\text{log}_{e}$.}
\begin{equation}\label{Dlog}
\begin{small}
\begin{split}
&D_{L}(z) =  \frac{\text{ln}(10)}{H_{0}}  \Bigg\{ \text{log}(1+z) + a_{2} \text{log}^{2}(1+z) + \\& + a_{3} \left.\Bigg[ k_{32} \text{log}^{2}(1+z) + \text{log}^{3}(1+z) \right.\Bigg] + \\& + a_{4} \left.\Bigg[ k_{42} \text{log}^{2}(1+z) + k_{43} \text{log}^{3}(1+z) +\text{log}^{4}(1+z) \right.\Bigg] + \\& + a_{5} \left.\Bigg[ k_{52} \text{log}^{2}(1+z) + k_{53} \text{log}^{3}(1+z) + k_{54} \text{log}^{4}(1+z) + \text{log}^{5}(1+z) \right.\Bigg] \Bigg\} .
\end{split}
\end{small}
\end{equation}
In our analysis, we tested the need for the different orders in this expansion proving the significance of the fifth-order (see Figure \ref{orthogonality} and the final comment of this section), while the sixth-order term turned out to be consistent with zero. This is the reason why we always consider an expansion to the fifth-order throughout the paper.\\
The terms $\displaystyle \frac{\text{ln}(10)}{H_{0}}$ and $a_{1}=1$ are needed to reproduce the Hubble law locally, that is $\displaystyle D_{L}(z) = \frac{\text{ln}(1+z)}{H_{0}}$. The coefficients $k_{ij}$ in this formula are not free parameters, and depend on the data set we want to fit and, in particular, on its redshift distribution. They are determined through the step-by-step procedure described in detail in Appendix \ref{appendixk}. Nevertheless, our results are not affected by applying different sub-selections for the quasar sample. This is mainly because the quality filters employed in the selection of quasars produce a homogeneous sample of quasars \citep{2020A&A...642A.150L} and, as a consequence, a change in the final cleaned sample does not modify significantly the results even if the coefficients $k_{ij}$ are formally sample-dependent.\\

\subsection{Data set and consequences of the "orthogonalization" procedure}
Our data set consists of the sample of SNe from \textit{Pantheon} survey \citep{scolnic2018} and the sample of quasars described in \citet{2020A&A...642A.150L} with the cutting at $z>0.7$ (see Section 8 of the paper for an explanation of this selection filter). For a detailed description and validation of the procedure used to turn quasars into standard candles and for the fitting technique used to include them in the cosmological analysis, we refer to \citet{rl19}, \citet{salvestrini2019}, \citet{lr16}, \citet{rl15}, \citet{2020A&A...642A.150L}. Here, we only point out that the non-linear relation between the X-ray and the UV emission of quasars, used to turn them into standard candles, depends on the luminosity distance and so it requires the assumption of a specific cosmological model. We will further discuss this point when we explain the mocking procedure applied on quasars in Section \ref{mocksamples}.\\

\begin{figure*}
    \centering
    \includegraphics[width=9.2cm]{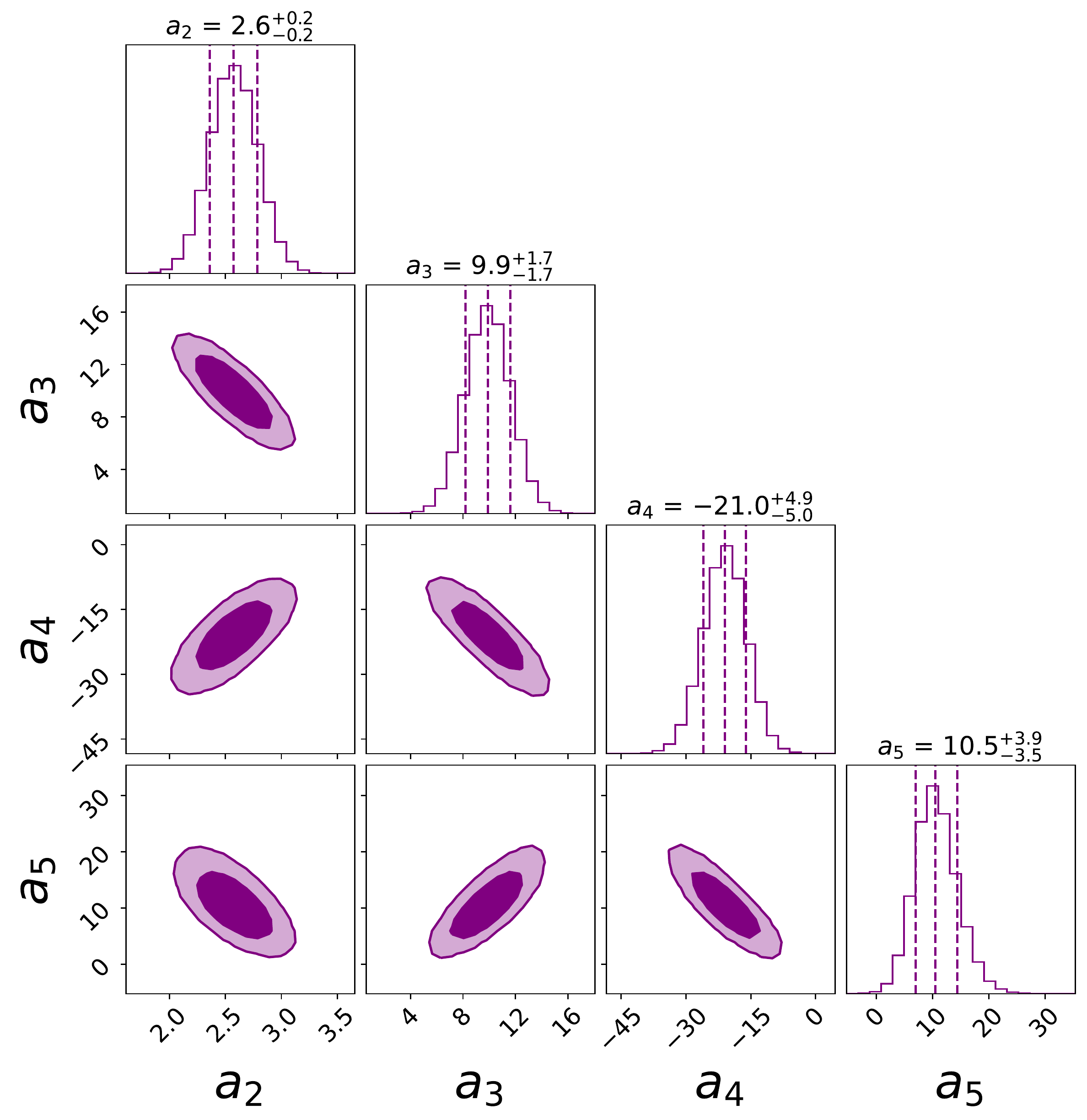}\includegraphics[width=9.2cm]{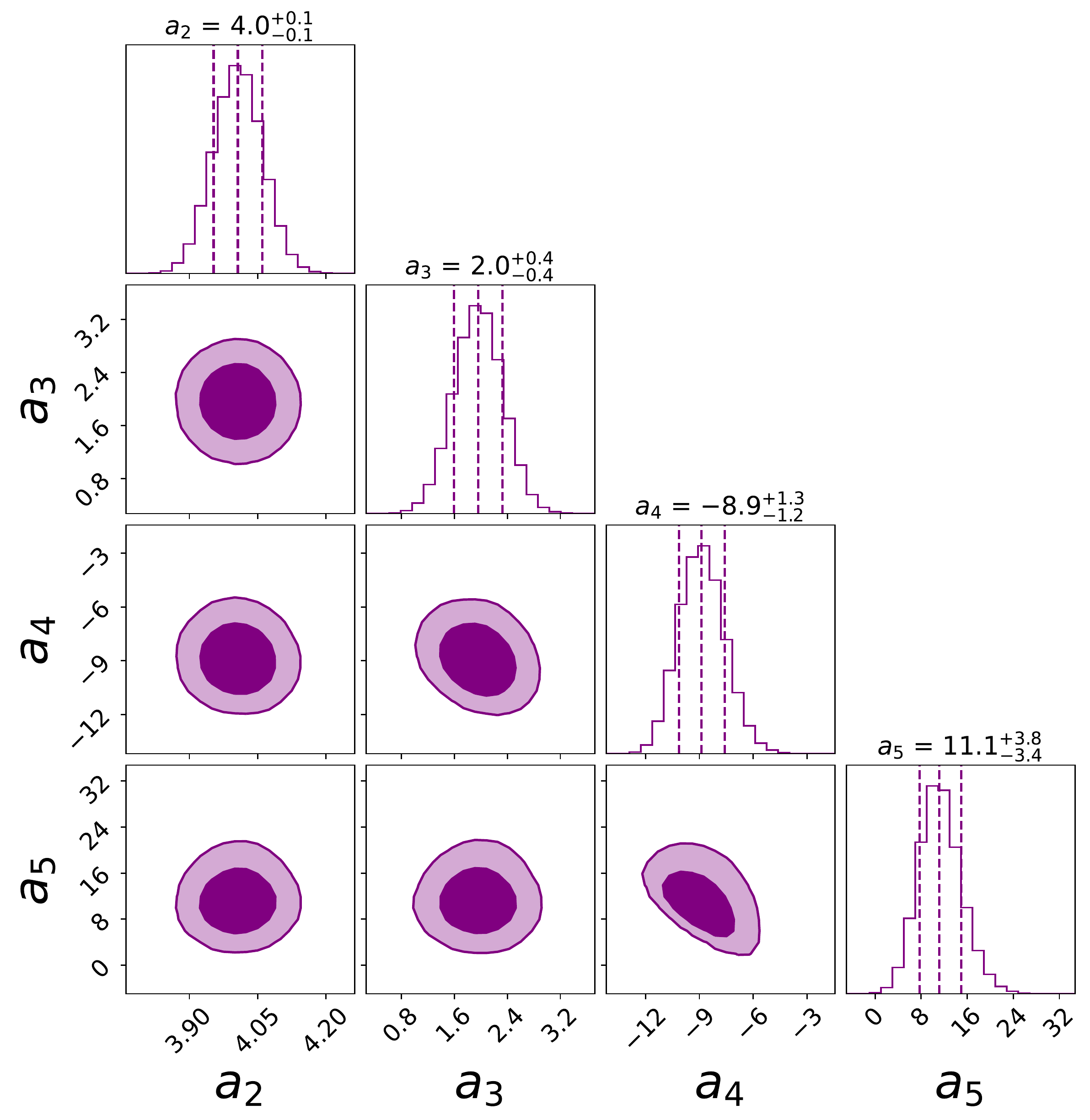}
    \caption{Contour plots of the cosmographic free parameters $a_{i}$ for the non-orthogonal (left panel) and orthogonal (right panel) logarithmic fits to the fifth-order. Purple contours are the confidence levels at 1 and 2$\sigma$. Best-fit values are very different in the two cases except for $a_{5}$, which is consistent within the errors. This is expected from equation \eqref{Dlog} as $a_{5}$ is not multiplied by any factor $k_{ij}$ and so it has to be equal to the same coefficient in the non-orthogonal formula. We stress that the apparent correlation in ($a_{3},a_{4}$) and ($a_{4},a_{5}$) planes in the right panel is due to the prior $D_{L}>0$ which allows only specific values of $a_{i}$.}
    \label{orthogonality}
\end{figure*}

\begin{figure}
    \resizebox{\hsize}{!}{\includegraphics{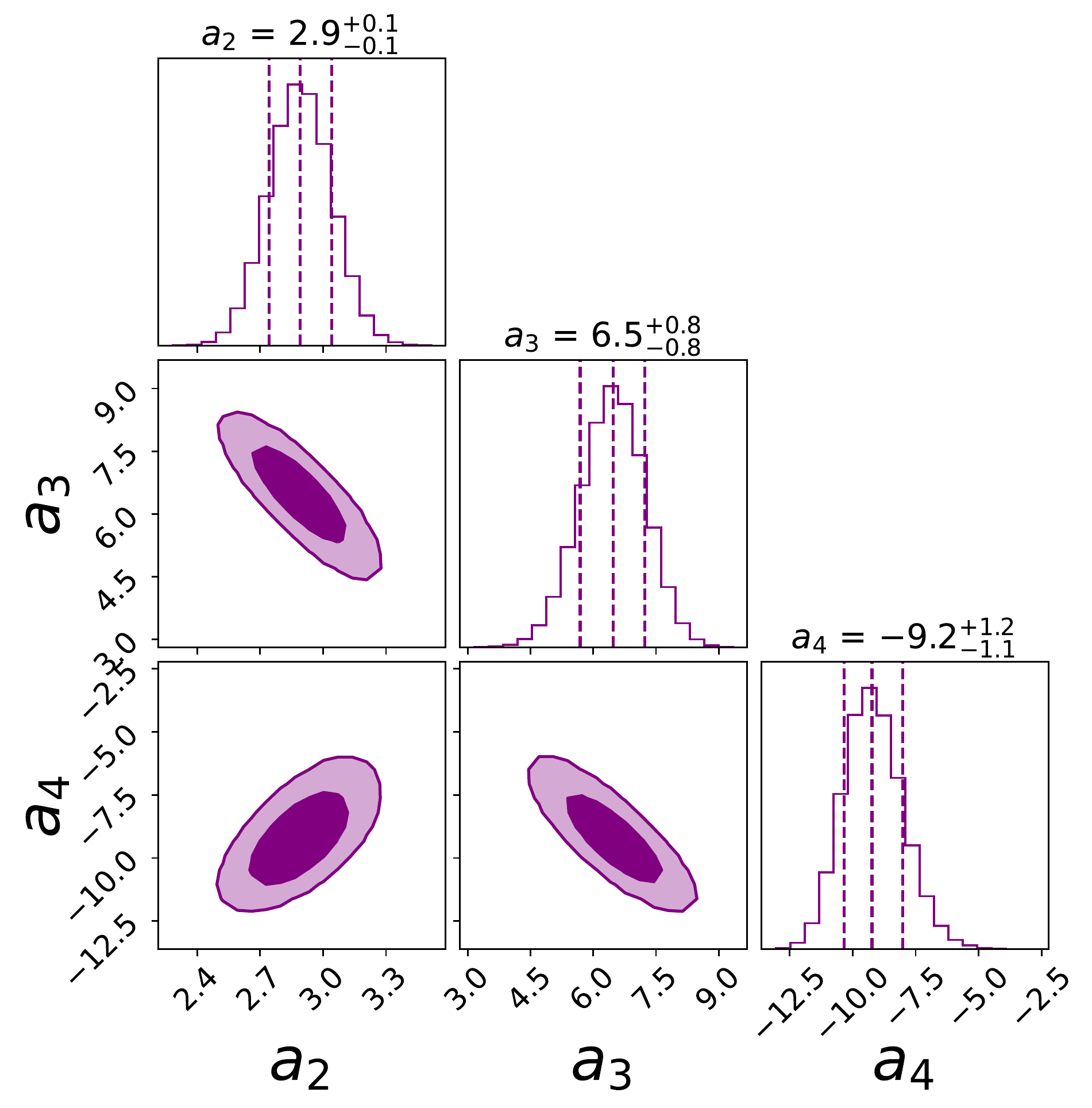}}
    \caption{Same as Figure \ref{orthogonality} but for the fourth-order non-orthogonal fit. We can notice that the best-fit values of $a_{i}$ are different from the ones of the fifth-order non-orthogonal fit showed in Figure \ref{orthogonality} as a consequence of the correlation between the parameters. These values are also inconsistent with the ones of the fifth-order orthogonal fit except for $a_{4}$, as already explained for $a_{5}$ in the caption of Figure \ref{orthogonality}.}
    \label{non-ortho}
\end{figure}

Figure \ref{orthogonality} shows the comparison between orthogonal and non-orthogonal\footnote{The formula for $D_{L}$ in the fifth-order non-orthogonal case is simply \begin{equation}\begin{split}\displaystyle &D_{L}(z)= \frac{\text{ln}(10)}{H_{0}} \Bigg[\text{log}(1+z) + a_{2} \text{log}^{2}(1+z) + a_{3}\text{log}^{3}(1+z)+ \\& + a_{4}\text{log}^{4}(1+z)+ a_{5} \text{log}^{5}(1+z) \Bigg].\end{split}\end{equation}} cosmographic logarithmic fits to our data set (quasars + SNe). In the latter case, as we can see from the slanting contours (at 1 and 2$\sigma$) in the parameter spaces of the left panel, the coefficients $a_{i}$ are correlated. As a consequence, a change in the order of the polynomial would modify the best-fit values of the parameters. We can directly see this comparing the non-orthogonal fourth-order fit shown in Figure \ref{non-ortho} to the one of the fifth-order in Figure \ref{orthogonality}. Contrarily, in the case of the orthogonal polynomial (right panel of Figure \ref{orthogonality}) there is no covariance at all between the parameters.\\

As anticipated, the results of the fits in Figure \ref{orthogonality} show that the fifth-order is significant, with a deviation from zero of about 3$\sigma$, and so it is required to fit our set of data.

\section{Validation of the cosmographic logarithmic approach}
\label{section3}

%
%
The cosmographic function is fitted over a redshift range much broader than the convergence radius of the logarithmic expansion. Therefore, we need to check if such expansion is still useful to test physical models. We note that in this paper we are only interested in the validation of the cosmographic approach within the redshift range of the observational data, while we do not discuss the extrapolation of the model to higher redshifts, and the related convergence issues. 

In order to compare a cosmographic fit with the predictions of a physical model, we use the relations that hold between the  cosmographic and physical free parameters, as described in Section \ref{section2}. 
However, this technique is strictly valid only in a redshift range up to $z \sim 1$ (due to the expansions centered in $z=0$). The comparison with a fit on the whole redshift range may not be strictly correct. Nevertheless, it could be possible that the cosmographic expansion represents a good approximation even if the redshift range goes beyond the convergence radius just because its shape is similar to the one of the cosmological model.\\
We investigated this issue in two ways: first, we compared the cosmographic expansion of the flat $\Lambda$CDM model with the model itself for different values of $\Omega_{M,0}$ (Section \ref{convergencetest}), then we compared the predictions of the cosmographic expansion with the fits of mock samples in the z=0-7.5 range (Section \ref{mocksamples}).

\subsection{Cosmographic expansion of physical models}
\label{convergencetest}

In order to check whether the cosmographic logarithmic expansion is a good approximation of the $\Lambda$CDM model beyond the convergence radius, we analyzed the ratio between the physical model and its logarithmic expansion, as a function of the redshift.
\begin{figure}
    \resizebox{\hsize}{!}{\includegraphics{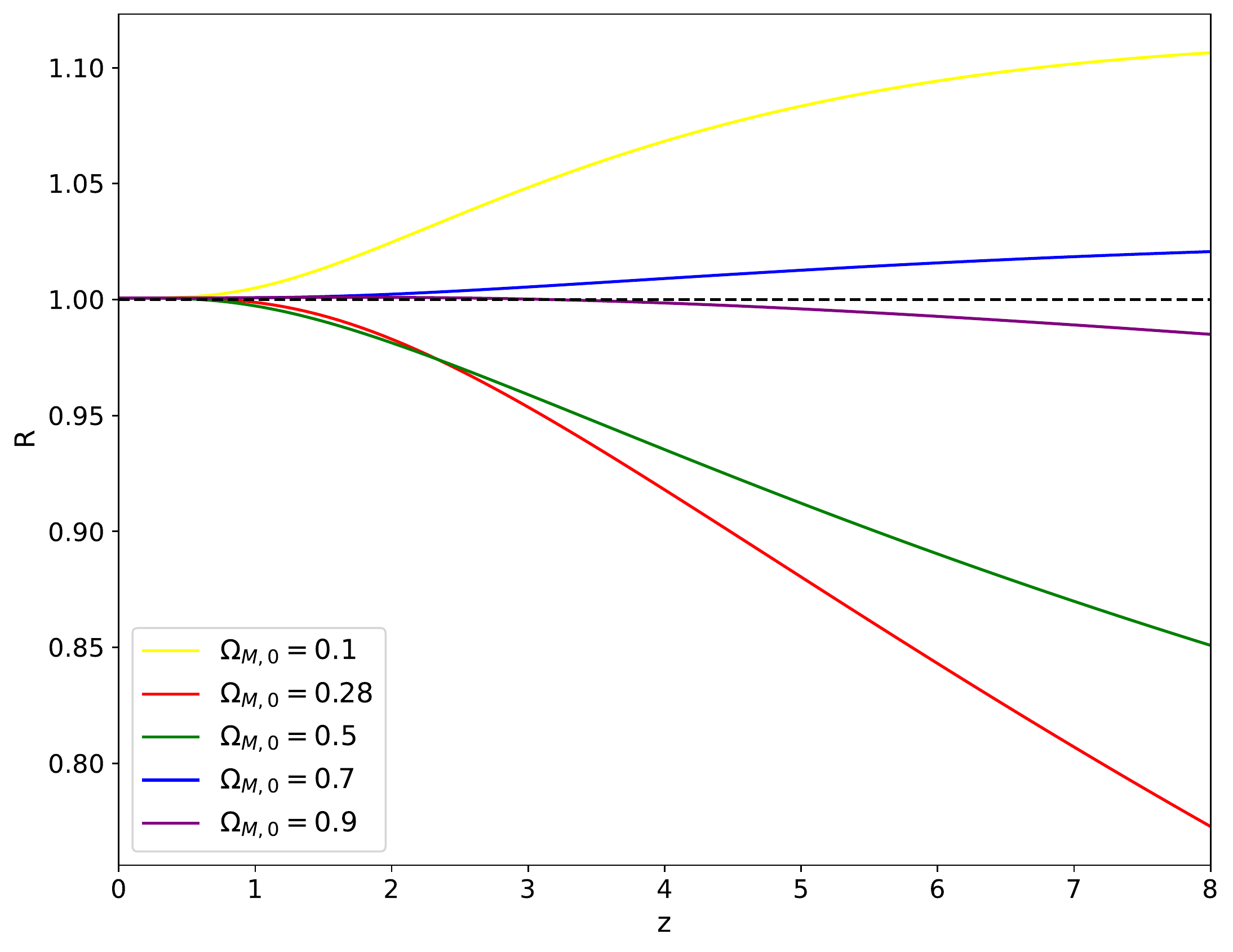}}
    \caption{Luminosity distance ratio ($R$) between a fifth-order cosmographic expansion of a flat \text{$\Lambda$}CDM model and the model itself, for different values of  $\Omega_{M,0}$. In each case the expansion very precisely reproduces the cosmological model for $z<1$ ($R=1$). The discrepancy from $R=1$ at $z>1$ depends on the specific value of $\Omega_{M,0}$: the worst case is the one with $\Omega_{M,0}=0.28$ while $\Omega_{M,0}=0.9$ is the best one.}
    \label{fig:convergenza}
\end{figure}
The results are shown in Figure \ref{fig:convergenza} (which reproduces Figure 1c of \citealt{2020arXiv200904109B}). The parameter $R$ represents the ratio between the luminosity distance estimated from the fifth-order cosmographic approximation of a specific flat \text{$\Lambda$}CDM model with $H_{0} = 70\,  \mathrm{km} \, \mathrm{s^{-1}} \,\mathrm{Mpc^{-1}}$ and five different choices of $\Omega_{M,0}$ ($\Omega_{M,0}=0.1,0.28,0.5,0.7,0.9$) and the luminosity distance from the corresponding cosmological model. The numerator of $R$ is obtained using the relations between $a_{i}$ and $\Omega_{M,0}$ calculated through the expansions in $z=0$ as described above. These relations depend both on the truncation order of the expansion and on the sample, due to the presence of the coefficients $k_{ij}$.  We refer to Appendix \ref{appendixcoeff} for the exact expressions of $a_{i}\,(\Omega_{M,0})$ in a logarithmic expansion at fourth and fifth-order. The redshift range we considered in Figure \ref{fig:convergenza} corresponds to the one spanned by our data, up to $z\simeq7.5$.\\
From this analysis we conclude that in all the cases the cosmological model is well reproduced up to $z \sim 1$. This implies that it is correct to compare the best-fit values of the cosmographic free parameters and the prediction of the cosmological model for these parameters only for fit at $z \leq1$. Moreover, the logarithmic expansion always shows a limited deviation from the \text{$\Lambda$}CDM model within the whole redshift range: the larger discrepancy from $R=1$ is $\sim 0.2$ at the maximum redshift $z \sim 7.5$ for the case with $\Omega_{M,0}=0.28$, which corresponds to a relative difference in distance modulus (DM) of 0.01. This implies that the logarithmic approximations and the reference cosmological models have the same "shapes", that is to say the same trend with redshift.\\

\subsection{Comparison with fits of mock samples}
\label{mocksamples}

\begin{figure*}
    \centering
    \includegraphics[width=6.2cm]{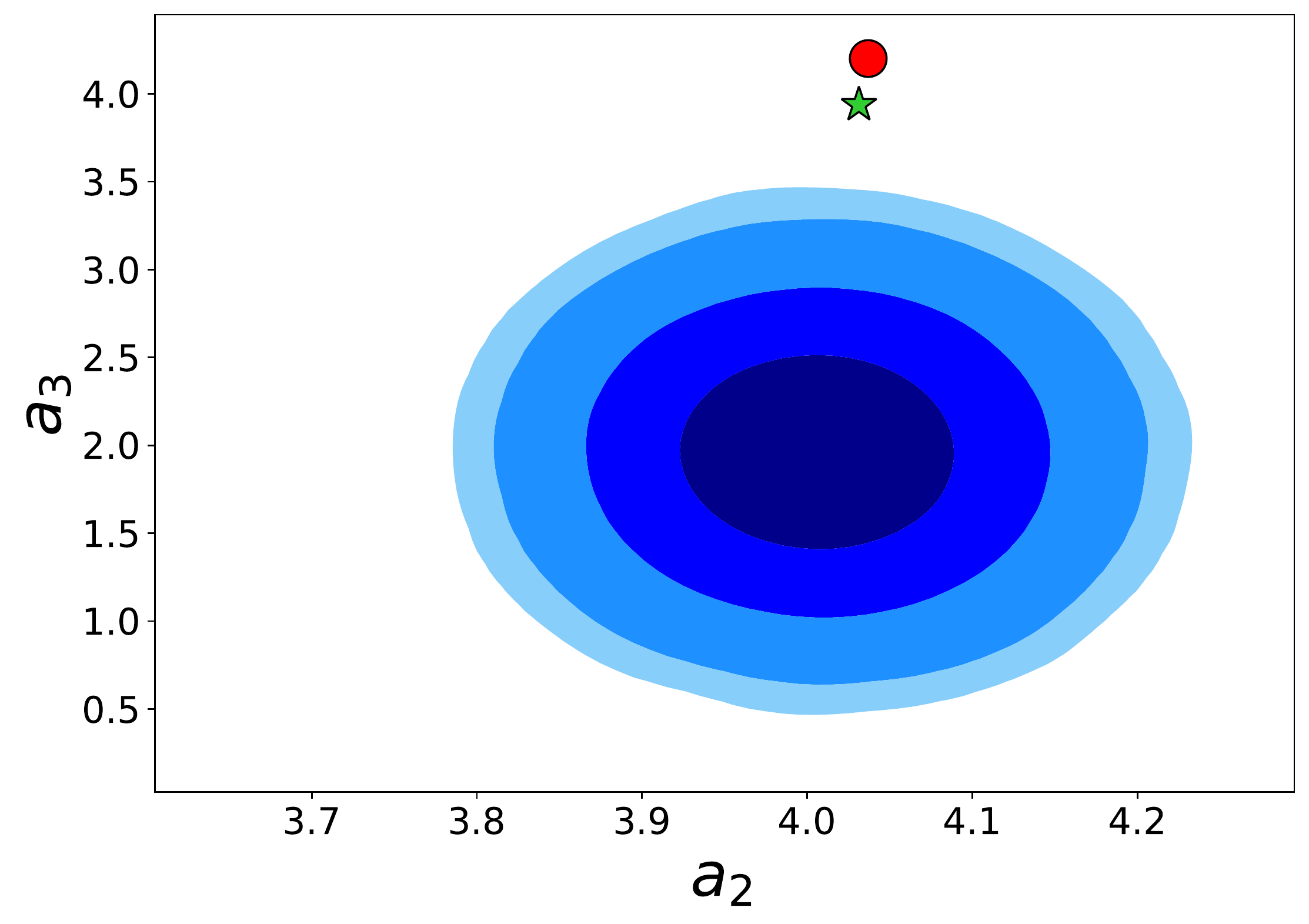}\includegraphics[width=6.2cm]{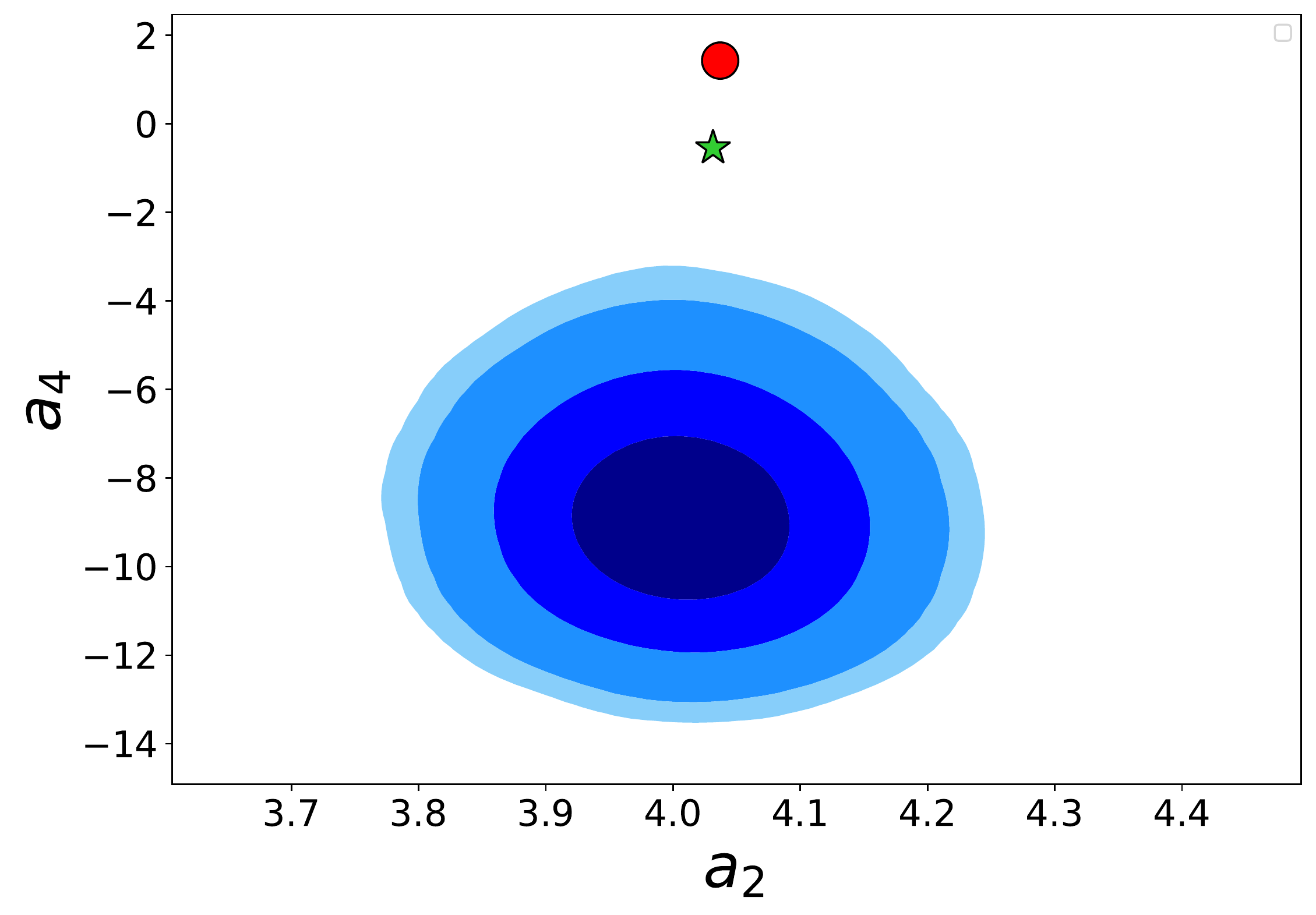}\includegraphics[width=6.2cm]{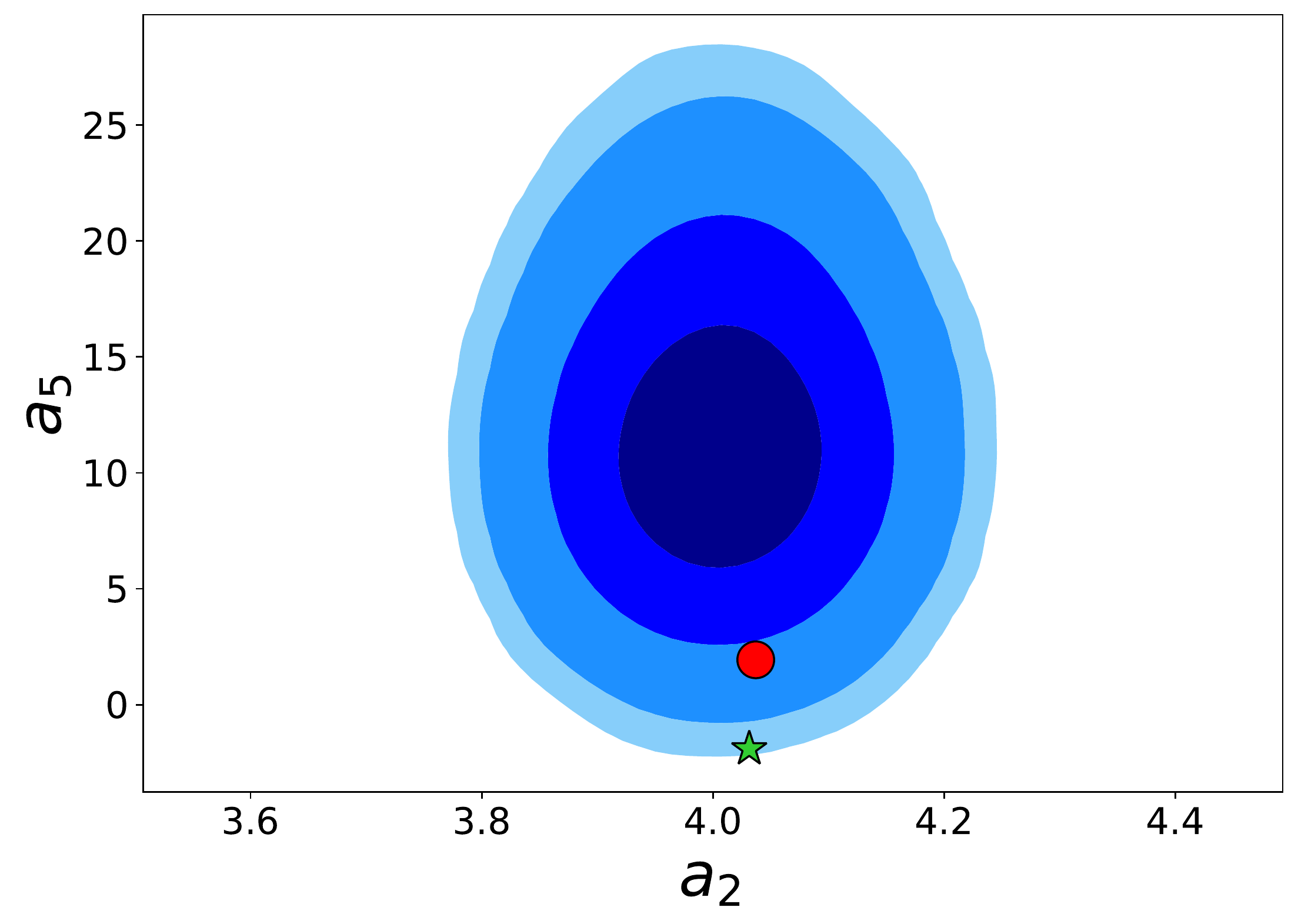}\\\includegraphics[width=6.2cm]{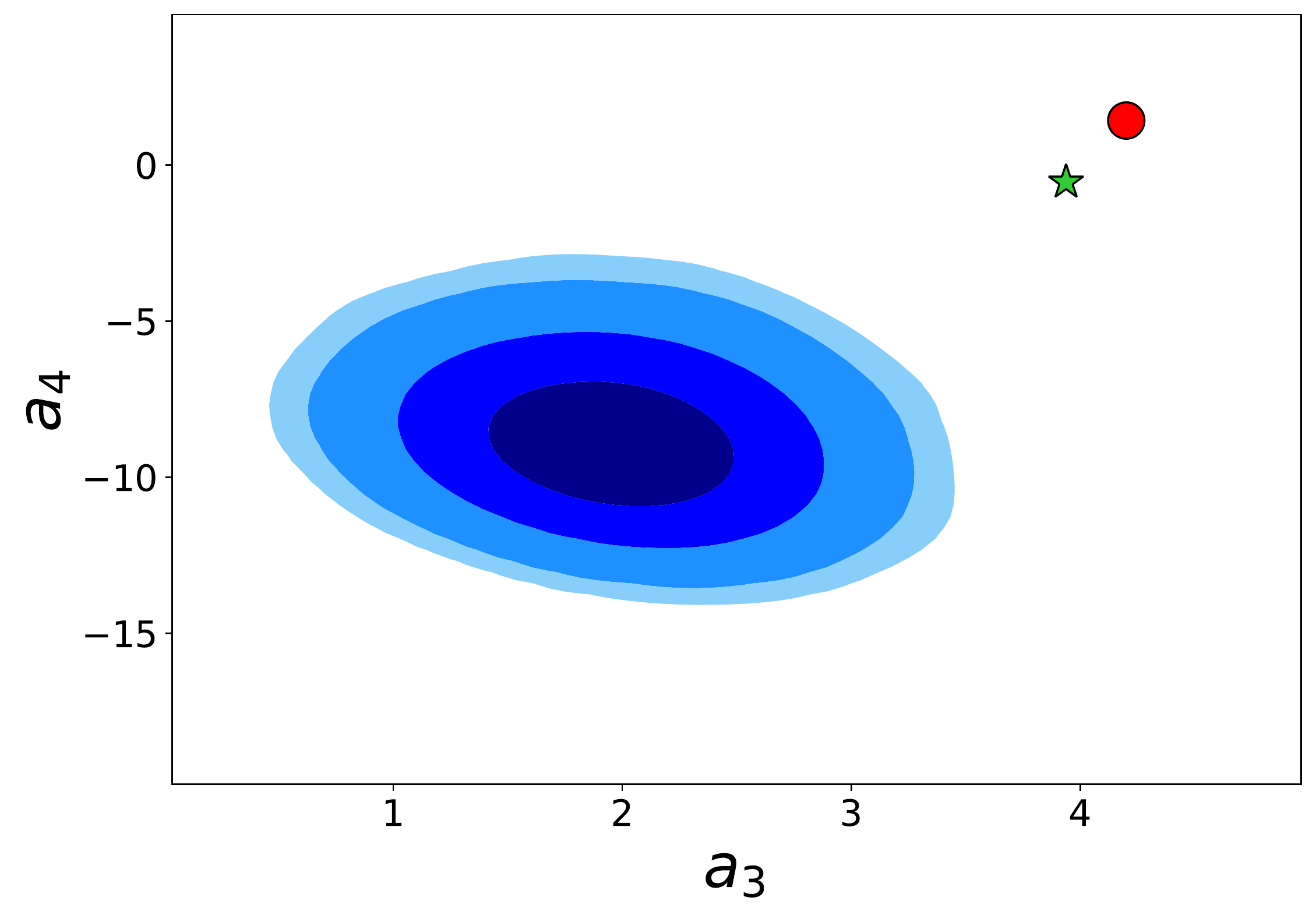}\includegraphics[width=6.2cm]{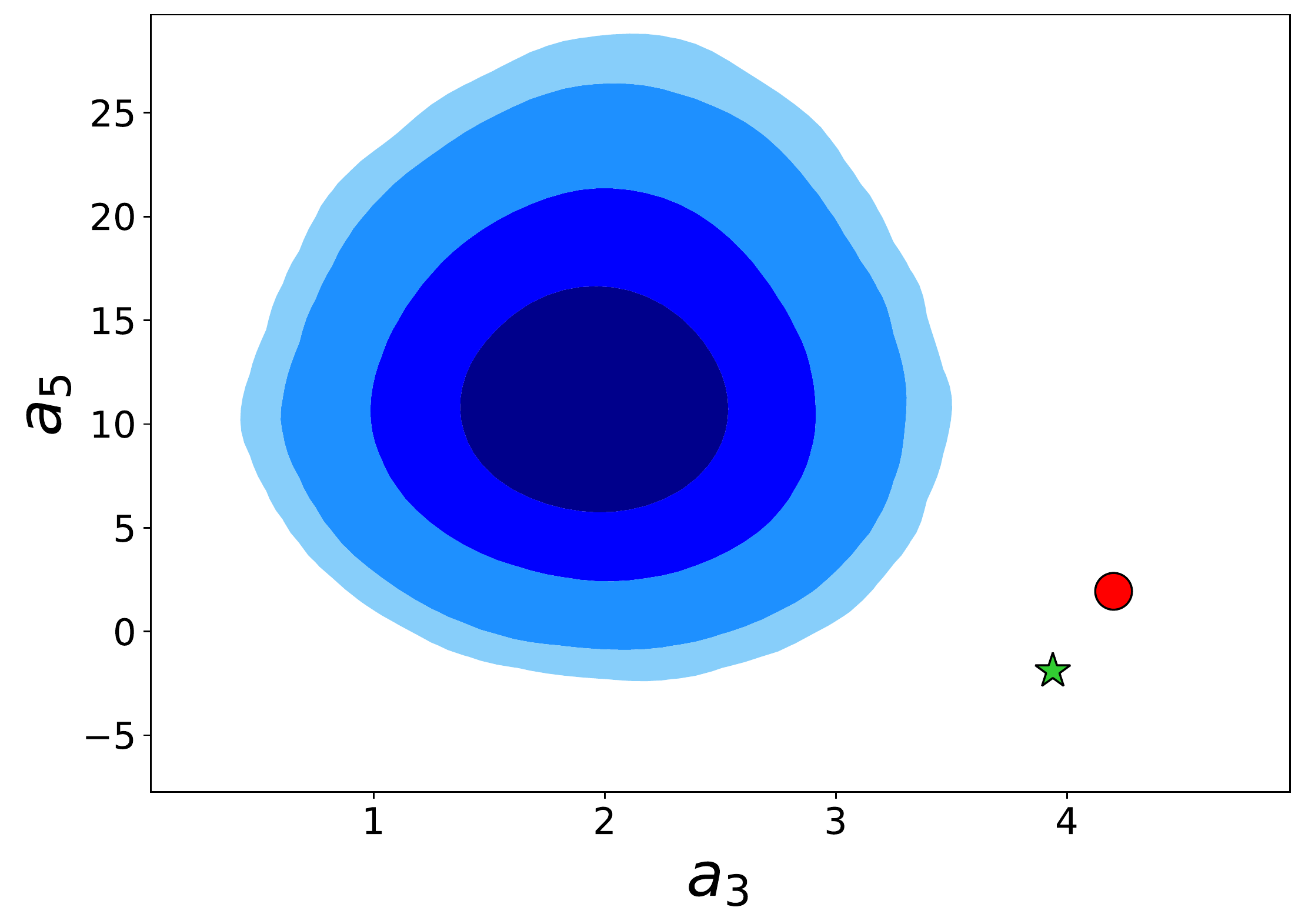}\includegraphics[width=6.2cm]{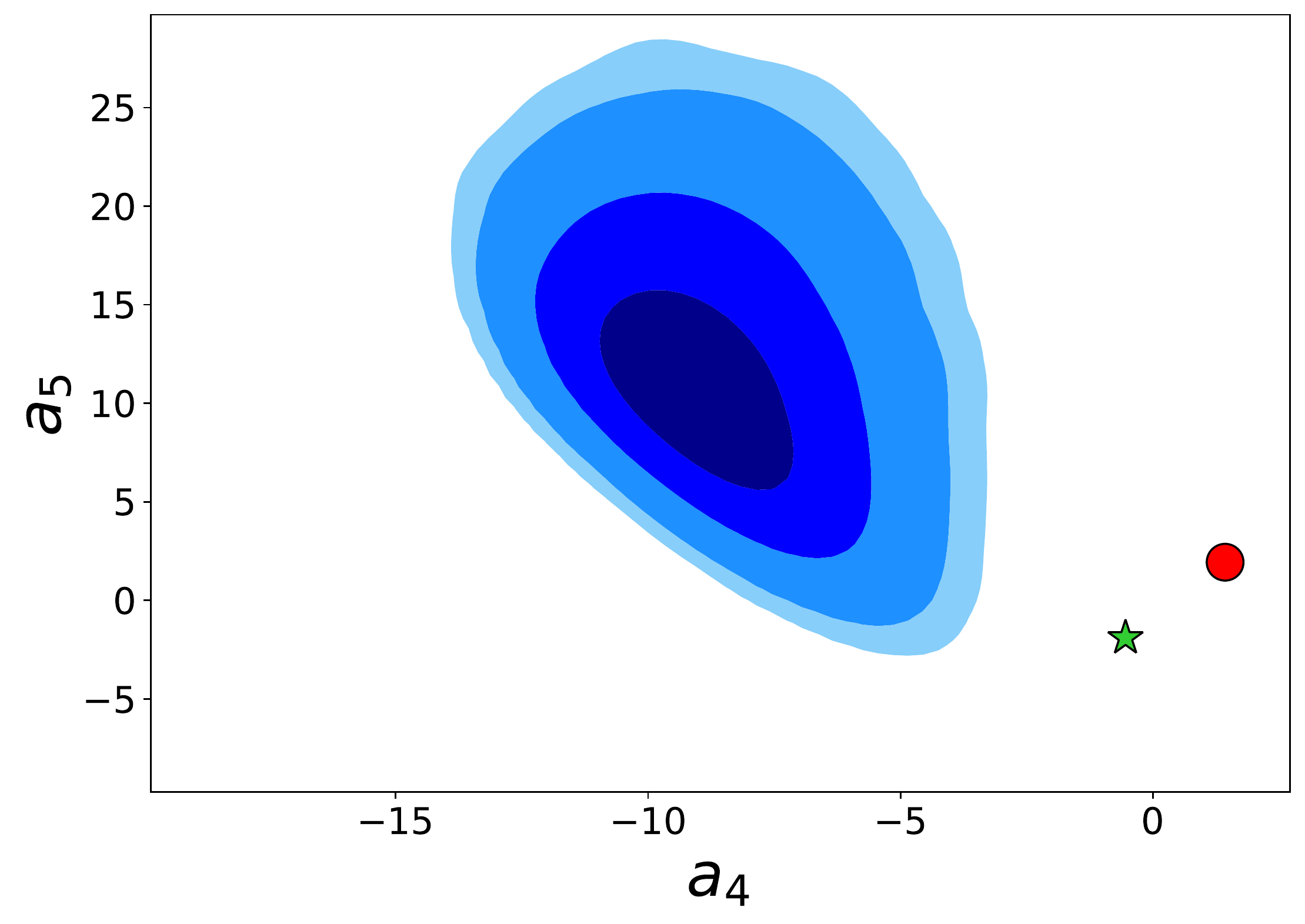}
    \caption{Bi-dimensional spaces for the cosmographic free parameters $a_{i}$. In each panel, the blue contours are the confidence levels at 1, 2, 3 and 4$\sigma$ for the best-fit on the real set of data (quasars+SNe), the red point is the best-fit from the mock sample with $\Omega_{M,0}=0.28$ and the green star corresponds to the prediction of the corresponding \text{$\Lambda$}CDM model from the expansion in $z=0$.} 
    \label{fig:mock0.28log}
\end{figure*}

\begin{figure*}
    \centering
    \includegraphics[width=6.2cm]{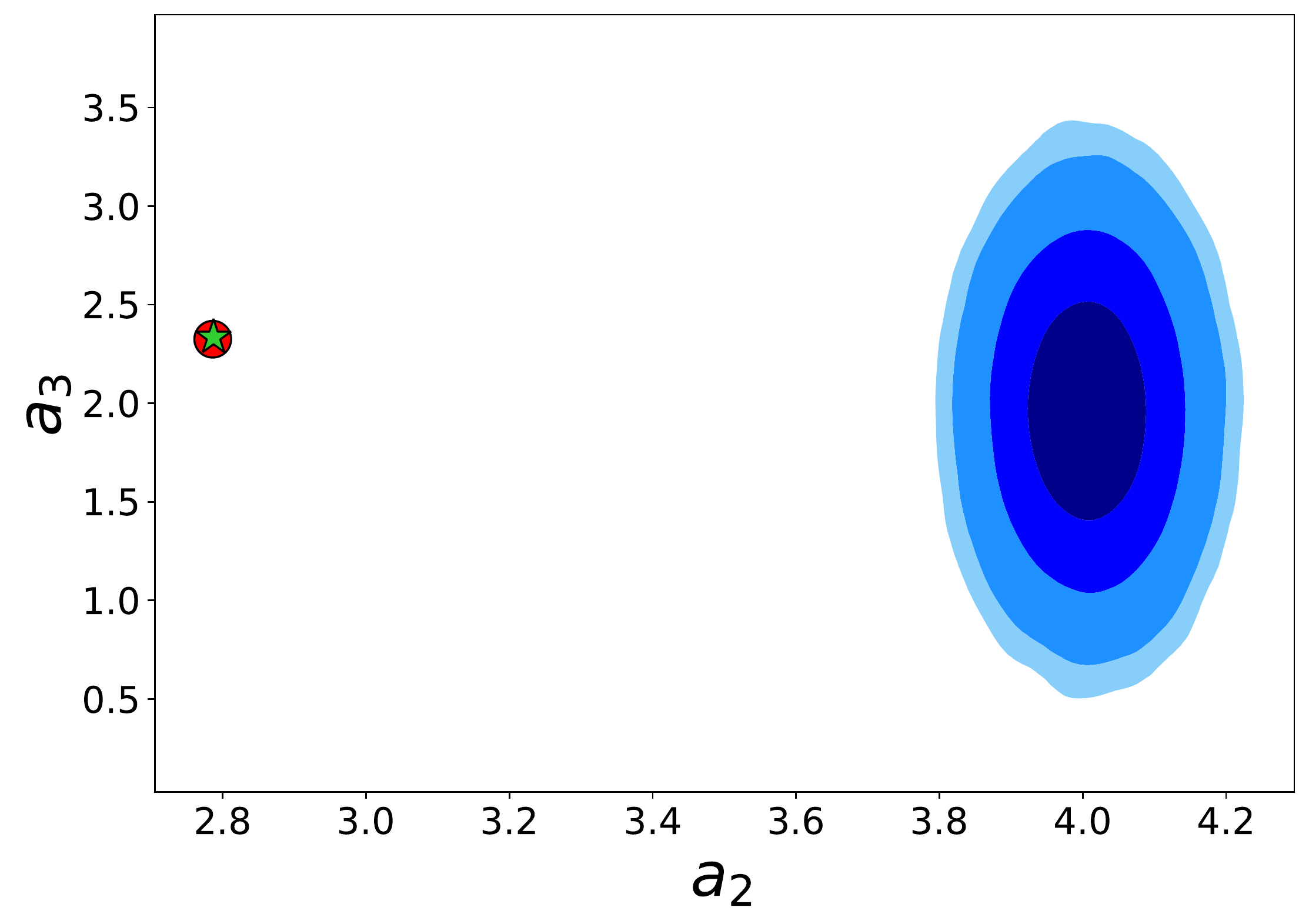}\includegraphics[width=6.2cm]{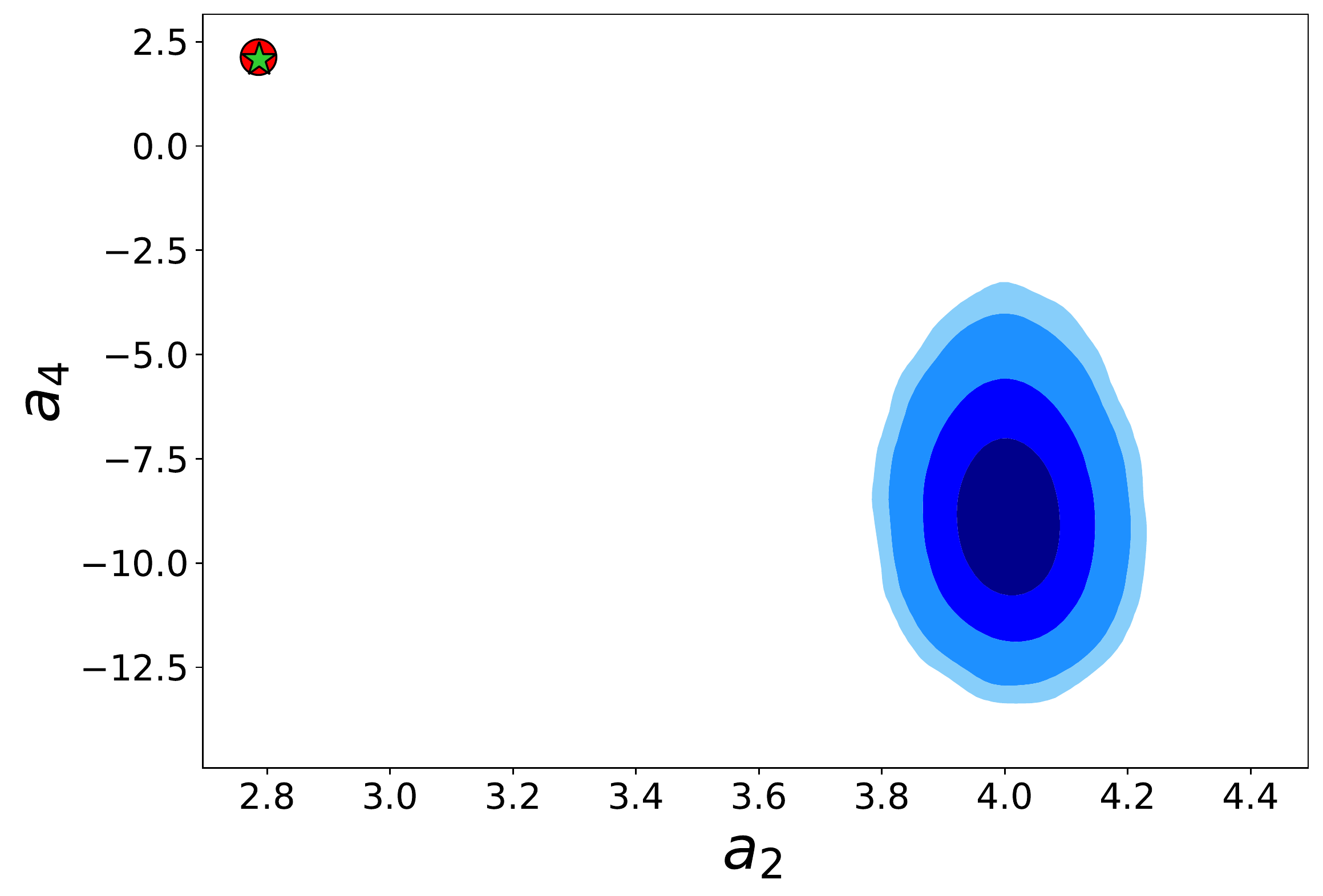}\includegraphics[width=6.2cm]{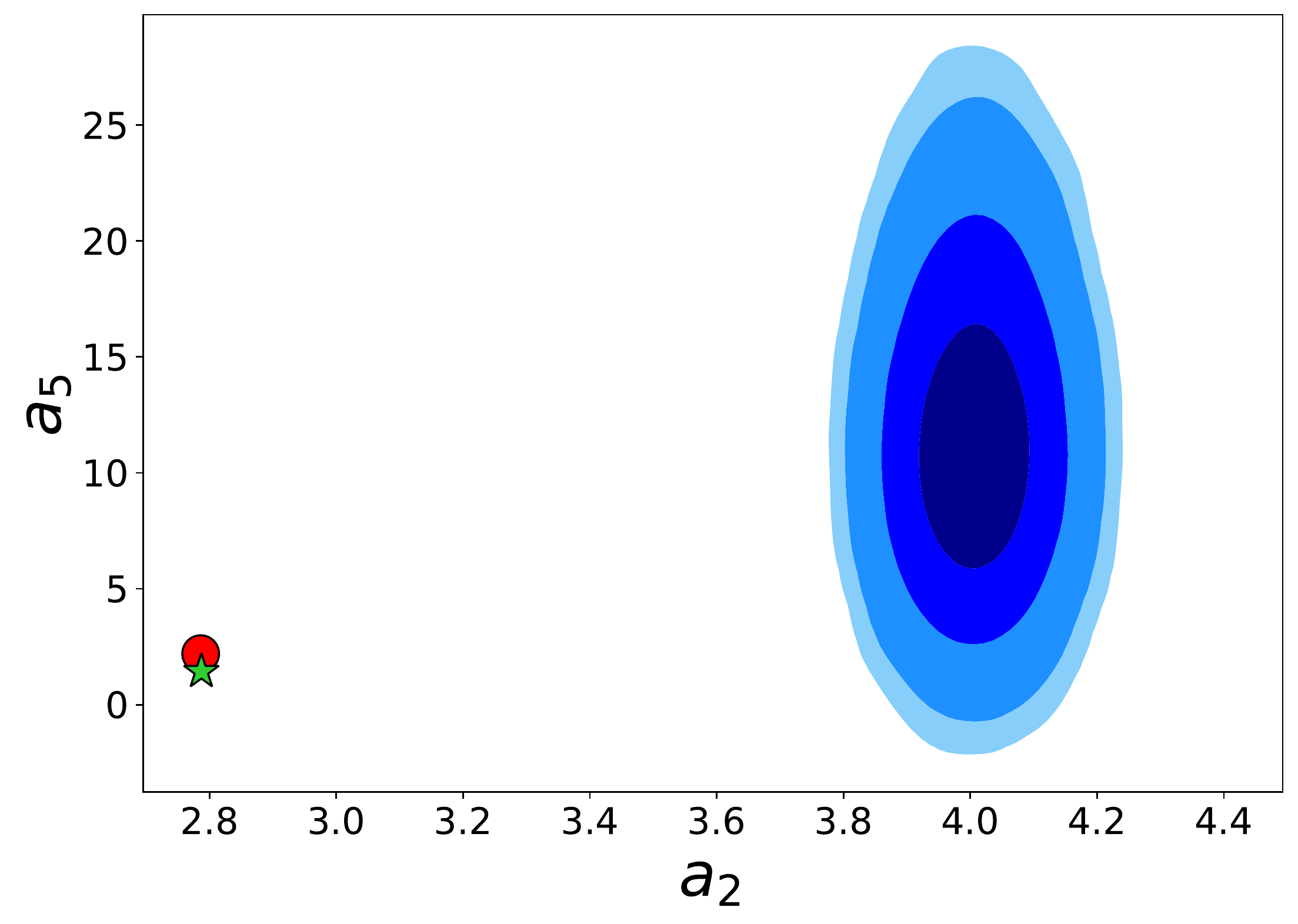}\\\includegraphics[width=6.2cm]{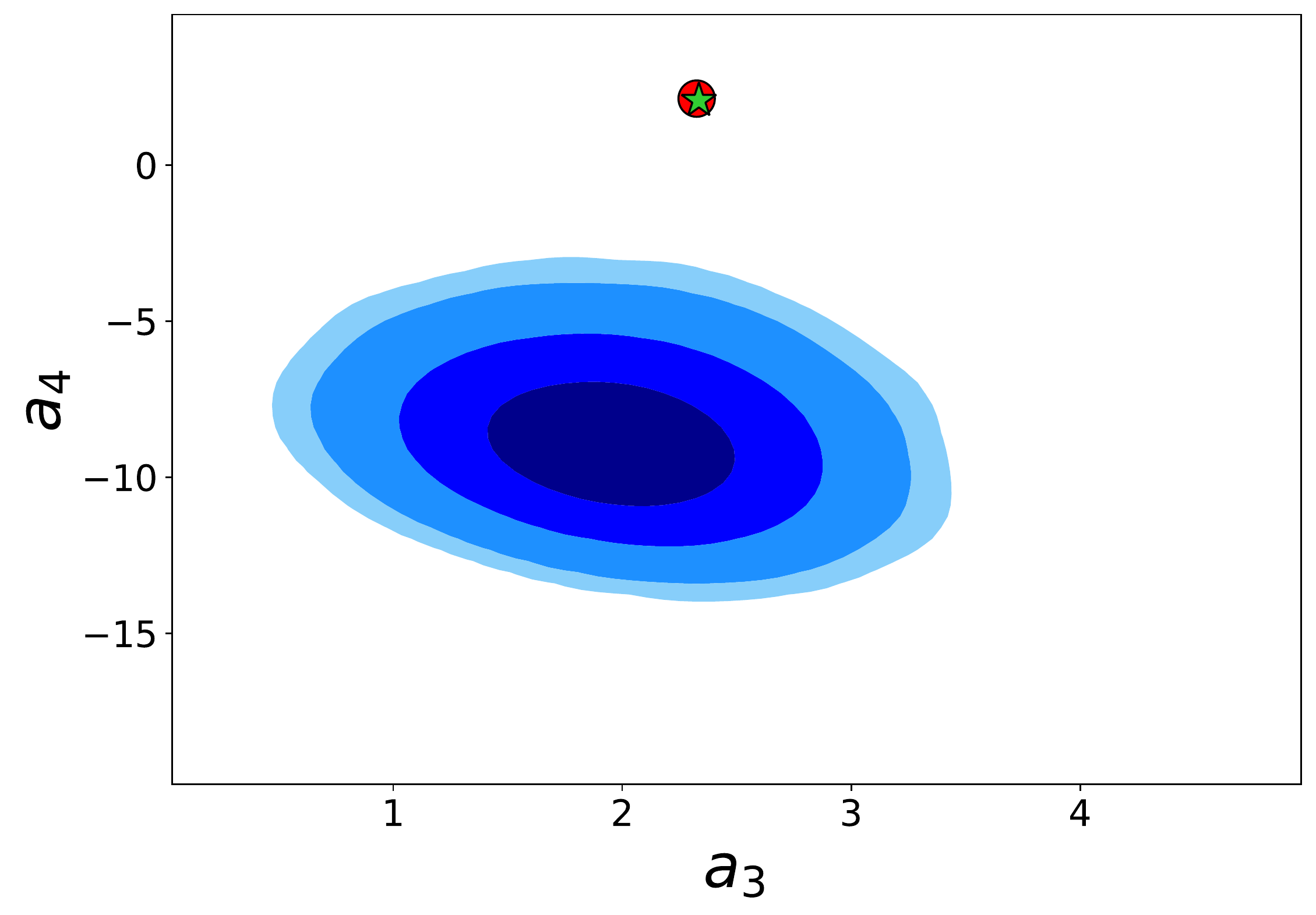}\includegraphics[width=6.2cm]{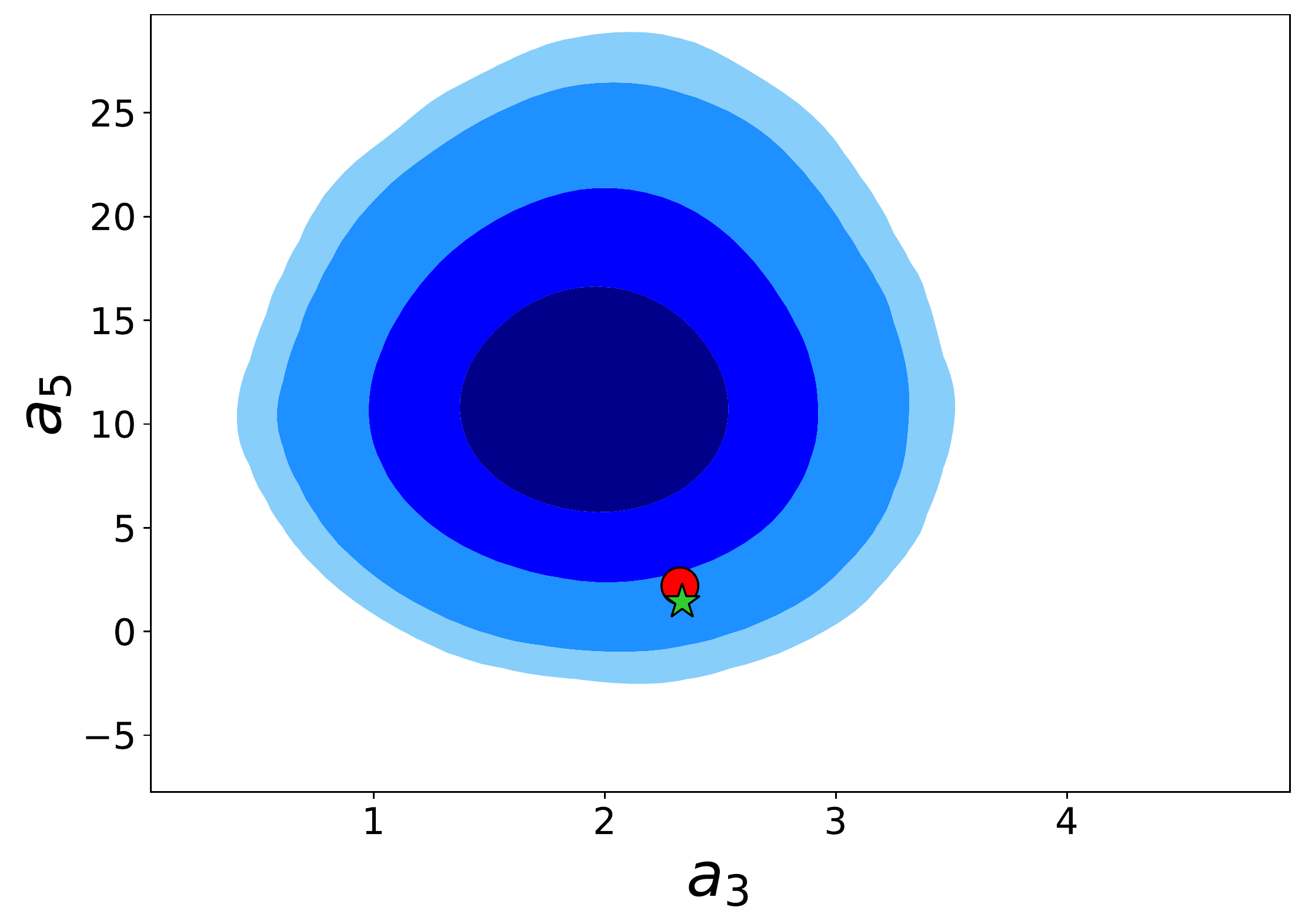}\includegraphics[width=6.2cm]{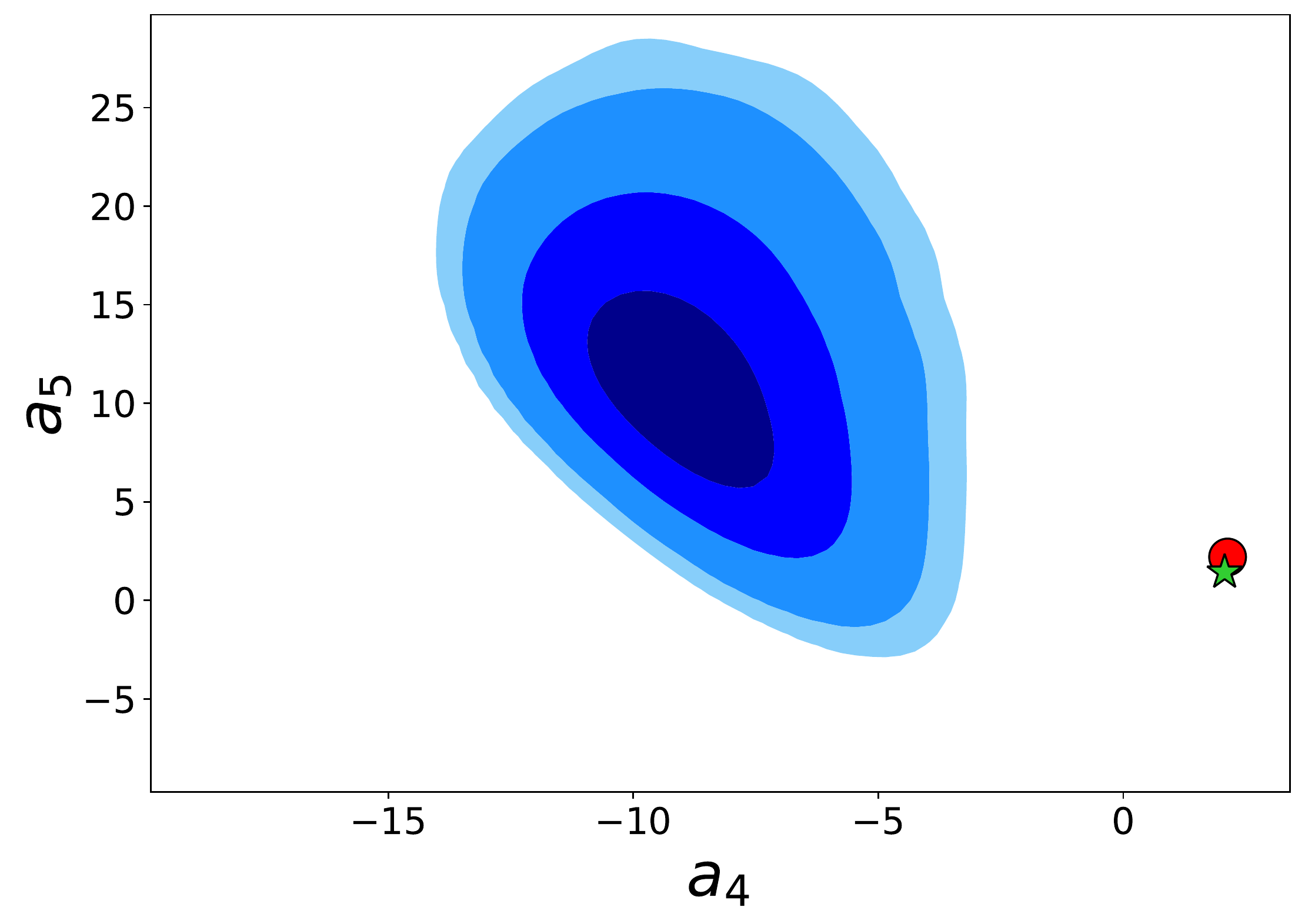}
    \caption{Same as Figure \ref{fig:mock0.28log} but with a reference flat \text{$\Lambda$}CDM model with $\Omega_{M,0}=0.7$.} 
    \label{fig:mock0.7log}
\end{figure*}

A direct way to check the applicability of a cosmographic analysis on a wide redshift range is through mock samples of data. Thanks to this procedure, we can simulate sets of data that have the same redshift distribution, statistical errors and intrinsic dispersion in distance modulus as the real sample and that follow a specific cosmological model. This sample is very different from the "ideal" one used in Figure \ref{fig:convergenza} in which statistical errors and dispersion are not taken into account and in which we have assumed an almost continuous distribution in redshift. As a consequence, a mock sample can be used to fit the cosmographic model on a "real" set of data that assumes a specific cosmological model and this fit can be compared with the theoretical prediction obtained by the analytical expansion around $z=0$. This approach allows to turn the qualitative comparison between $D_{L}(z)$ (as in Figure \ref{fig:convergenza}) into a quantitatively comparison in the $a_{i}$ bi-dimensional spaces. This is the crucial test to investigate if the prediction from an expansion at $z=0$ can be extrapolated up to redshifts $z \sim 8$.\\
We now explain in detail the procedure we followed starting from the creation of the mock samples of quasars and SNe.\\
Regarding SNe, we calculated the mock distance moduli using the redshifts from the \textit{Pantheon} sample and $D_{L}(z)$ from a flat \text{$\Lambda$}CDM model ($\text{DM}(z) = 5\, \text{log}(D_{L}(z)(Mpc)) + 25$) with $H_{0} = 70\,  \mathrm{km} \, \mathrm{s^{-1}} \,\mathrm{Mpc^{-1}}$ and two values of $\Omega_{M,0}$: $\Omega_{M,0} = 0.28$ and $\Omega_{M,0} = 0.7$. We also included in the calculation of DM the contribution of a Gaussian distribution with an intrinsic dispersion around the central value of 0.16 dex. The statistical errors associated with DM are the ones from the real sample. The mock samples of quasars have the same redshifts and UV luminosity as the real set of data \citep{2020A&A...642A.150L} and X-ray fluxes obtained from the X-UV relation under the assumption of a flat \text{$\Lambda$}CDM model with the same values of $H_{0}$ and $\Omega_{M,0}$ assumed for the mock samples of SNe. As for SNe, we took into account also their intrinsic dispersion. Specifically, we assumed a Gaussian distribution in DM around 1.3 dex. Once again the statistical errors on the distance moduli are the ones from the real sample of data. The number of sources in these mock samples is increased (for both quasars and SNe) in order to constrain the free parameters of the fits with higher precision leaving the relative statistical weight of quasars and SNe unchanged. For every redshift we generated 100 sources with slightly different DM due to the Gaussian distribution of the intrinsic dispersion. Thanks to this, we improved the precision on the best-fit of the free parameters of a factor 10 making the uncertainties negligible compared to the ones of real data\footnote{As a consequence the error bars are too small to be plotted on the red point in Figures \ref{fig:mock0.28log} and \ref{fig:mock0.7log}.}.\\
Following this method, we built two joint mock samples of quasars and SNe assuming $\Omega_{M,0}=0.28$ and $\Omega_{M,0}=0.7$, respectively. These choices are justified by the fact that, looking at Figure \ref{fig:convergenza}, these are the cases in which the cosmological model is worst and best (just after $\Omega_{M,0}=0.9$) reproduced by the expansion around $z=0$, respectively. The conclusions we obtain for these two extreme cases can be generalized to all the other intermediate ones. Moreover, $\Omega_{M,0}=0.28$ is the value actually favoured by several cosmological probes and so it is of common interest to test our method on this specific value.\\

As self-consistency check, we firstly recover the cosmological parameters through a direct fit of a flat \text{$\Lambda$}CDM model to the mock data. The inferred best-fit values of $\Omega_{M,0}$ and $H_{0}$ are shown in Table \ref{tab:cosmopar}. This test quantifies the degree of randomness introduced in the mocking procedure and confirms that the data are fully consistent with the specific flat \text{$\Lambda$}CDM model assumed.\\

\begin{table}[h!]
\centering 
\begin{tabular}{ |c|c|c| } 
 \hline
 mock data & $\Omega_{M,0}$ & $H_{0}\, (\mathrm{km} \, \mathrm{s^{-1}} \,\mathrm{Mpc^{-1}})$\\ 
 \hline
 $\Omega_{M,0}=0.28$ & $0.280 \pm 0.001$ & $70.00 ^{+0.02}_{-0.02}$ \\ 
 \hline
 $\Omega_{M,0}=0.7$ & $0.701 \pm 0.002$ & $70.02 ^{+0.02}_{-0.02}$ \\ 
 \hline
\end{tabular}
\caption{Best-fit values of $\Omega_{M,0}$ and $H_{0}$ from the mock samples of data built according to the correspondent cosmological models. The input parameters are perfectly recovered.}
\label{tab:cosmopar}
\end{table}
The mock samples of quasars and SNe obtained in this way are fitted jointly with a fifth-order orthogonal logarithmic function. We study the two cases of $\Omega_{M,0}=0.28$ and $\Omega_{M,0}=0.7$ separately finding, in the end, some generalized conclusions.\\
Figure \ref{fig:mock0.28log} summarises the results for the case with $\Omega_{M,0}=0.28$. Each panel in this figure shows a bi-dimensional free parameter space of $a_{i}$ in which: the confidence levels from 1 to 4$\sigma$ refer to the cosmographic analysis on the real set of data of quasars from \citet{2020A&A...642A.150L} and SNe from \textit{Pantheon} survey (these are exactly the ones already shown in the right side of Figure \ref{orthogonality}), the red point is the best-fit from the mock sample while the green star is the analytic prediction from the expansion of the standard model around $z=0$ (equations \ref{coefflcdm} for $\Omega_{M,0}=0.28$).\\
Focusing on the comparison between the red point and the green star, we see that they perfectly match on the parameter $a_{2}$, while the green star underestimates the discrepancy from observational data of 0.7$\sigma$ in $a_{3}$ and 1.5$\sigma$ in $a_{4}$ and overestimates the same discrepancy of 1$\sigma$ in $a_{5}$. This means that the prediction in the limit $z<1$ is in good agreement with the correct results also if evaluated on the whole redshift range. Furthermore, this analysis shows something more interesting under a deeper insight. The discrepancy between the data and the \text{$\Lambda$}CDM model with $\Omega_{M,0}=0.28$ (green star) is at more than 6/7$\sigma$ for $a_{3}$ and $a_{4}$, respectively. Moreover, for these two parameters, the prediction of the expansion at $z=0$ (green star) affects the results with an underestimation of the discrepancy if compared to the fit on the mock sample (red point). As a consequence, the claim for the existence of a tension coming from the comparison between data and analytic prediction would be correct and confirmed at even more sigmas by the use of mock samples. The results on the parameter $a_{5}$, instead, require a more complex interpretation. In this case, the discrepancy of the data with the red point (the best fit from the mock sample) and the green star (the prediction from the expansion) is at 2.7$\sigma$ and 3.8$\sigma$, respectively. Nevertheless, we need to consider that $a_{5}$ is the less significant parameter and the one that is less constrained because of the high errors. So we expect that, if a larger sample of quasars at high redshifts (where this term is dominant) was available, $a_{5}$ could be constrained better leading to a much more significant discrepancy from the cosmological model, such as the ones observed in $a_{3}$ and $a_{4}$. If this happens, the argument on $a_{3}$ and $a_{4}$ could be generalized also to this parameter. From this last point it is evident the crucial role of future observations of quasars at high redshifts, which could certainly make this analysis easier and clearer.\\
The complexity of this analysis and the care needed to take every aspects into account lead us, in the end, to demonstrate the great advantages of the logarithmic approach. On one side, the fifth-order orthogonal logarithmic function can precisely reproduce the given physical model, as proved by the fit on the mock data. On the other side, as we are only interested in comparing the data with cosmological models to test for a possible tension, we can directly compare the fit in the whole redshift range with the prediction from the expansion at $z=0$, without the need of mock samples, as the approximation introduced with this method is negligible.\\
We demonstrated it for $\Omega_{M,0}=0.28$, but, as already pointed out, this is the case in which the analytic prediction has the maximum deviation from the model (Figure \ref{fig:convergenza}) and so these conclusions can be generalized to all the other flat \text{$\Lambda$}CDM models with values of $\Omega_{M,0}$ between 0.1 and 0.9. Moreover, in all the other cases we expect a better agreement of the theoretical prediction and a solid confirmation of the results described above.\\
As an example of this statement, we report in Figure \ref{fig:mock0.7log} the same analysis for a flat \text{$\Lambda$}CDM model with $\Omega_{M,0}=0.7$. The key result is that the green star and the red point completely overlap in each of the parameter spaces leading to a perfect match between the physical model and the prediction at $z=0$.\\
This is a strong evidence that the cosmographic extrapolation can be used at redshifts $z>1$. This allows us to avoid fitting mock samples for every possible value of $\Omega_{M,0}$ to reproduce the curve predicted by a flat \text{$\Lambda$}CDM model in the parameter spaces substituting this long-time computational procedure with the simple analytic calculation at $z\sim0$.\\

Summarizing, we have studied in detail the features of our cosmographic approach showing that the logarithmic function has excellent properties and can reproduce physical models over the whole redshift range of the data considered. As a consequence, we can use this method to describe the Hubble diagram of quasars and SNe analytically and compare it with the predictions of cosmological models.\\
The method we have just described is completely general and can be applied to any cosmographic function to test if its theoretical prediction at $z=0$ can be compared to the cosmographic fit on the whole redshift range. This gives us an unbiased measure of the usefulness and the power of different cosmographic approaches.

\section{Test of the flat \text{$\Lambda$}CDM model}
\label{section4}

Once we have demonstrated the robustness of the cosmographic logarithmic technique, the next step is to use it as a test for cosmological models. We have proved that our cosmographic function gives a reliable reproduction of a flat \text{$\Lambda$}CDM model if we assume this model as the reference one, but we can also compare the prediction of the standard cosmological model with the observational data taking advantage of the degrees of freedom and the flexibility of the logarithmic function. This means fitting the data with both cosmological and cosmographic models with the purpose of studying: 1) the deviations of the distance moduli; 2) the different trend of the two best-fit curves in the Hubble diagram; 3) the discrepancy between the best-fit values of $a_{i}$ and the values of $a_{i}$ predicted by the cosmological model. Regarding this last point, we have already shown in the previous section that it is correct to compare the findings obtained with the data over the whole redshift range with the theoretically predicted values of $a_{i}(\Omega_{M,0})$ obtained by the expansion around $z=0$.\\
An analysis of this kind is possible only because we have defined a model that has good convergence properties and that guarantees a purely analytical reconstruction of the data as it does not rely on any physical assumptions on the evolution of the Universe or on the nature of its constituents (except for the assumption of the FLRW metric). In principle, this cosmographic model can be used as a test for any cosmological model, but in this paper, we focus only on a flat \text{$\Lambda$}CDM model, while the study of other cosmological models will be developed in forthcoming dedicated publications.\\

As the first step, we fitted the distance moduli of the joint sample of quasars \citep{2020A&A...642A.150L} and SNe Ia \citep{scolnic2018} with the cosmographic logarithmic orthogonal expansion to the fifth-order (equation \ref{Dlog} with the coefficients $k_{ij}$ fixed for this sample through the procedure described in Section \ref{appendixk}) and with a flat \text{$\Lambda$}CDM model. The fit of the flat \text{$\Lambda$}CDM model is performed not only on the whole redshift range of the data but also in the restricted interval with $z < 1.1$, where SNe are predominant, to get a deeper insight of the comparison between this cosmological model and the observational data.\\
\begin{figure}
    \resizebox{\hsize}{!}{\includegraphics{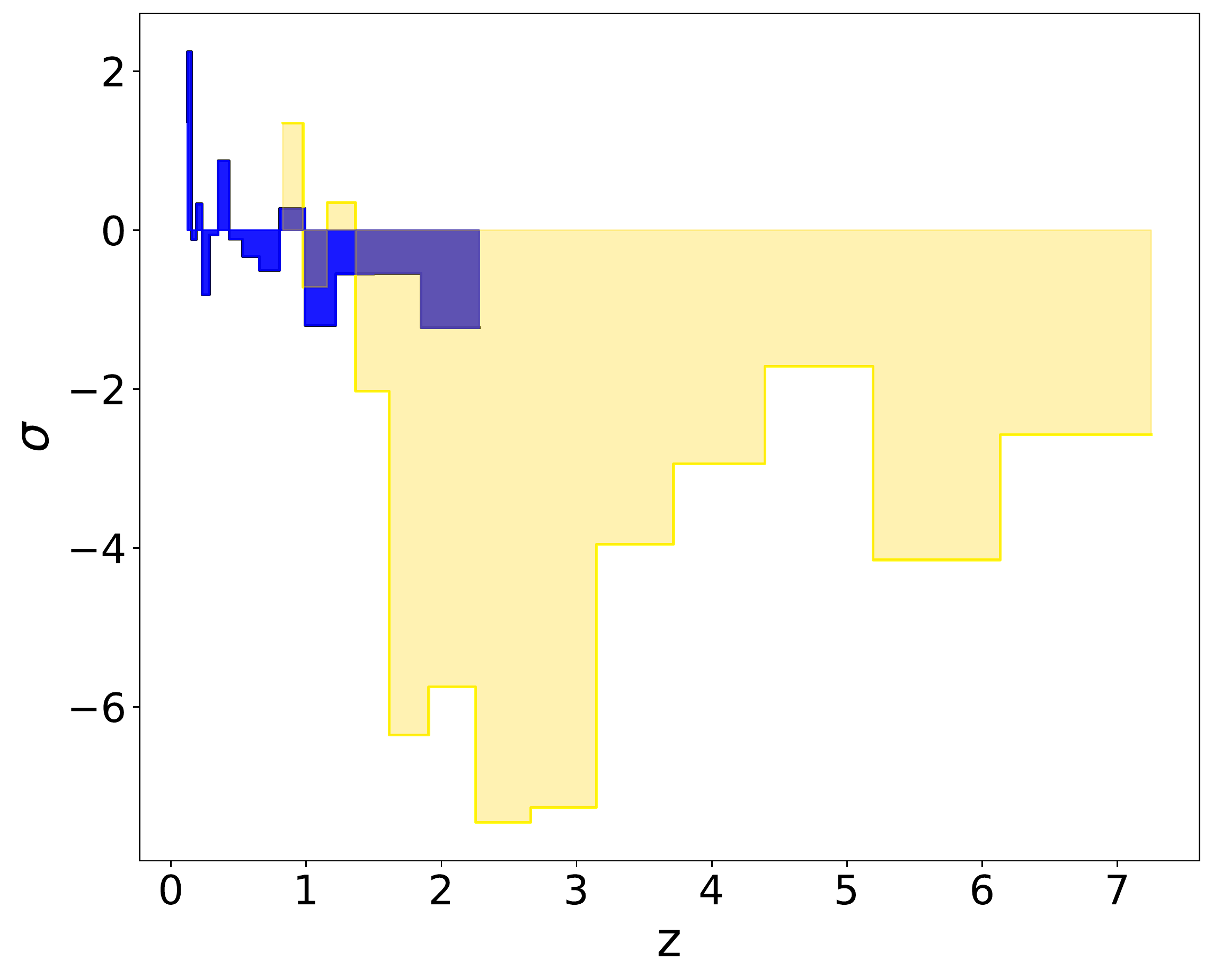}}
    \caption{Deviations (in unity of $\sigma$) of the cosmographic distance moduli of quasars (yellow bars) and SNe (blue bars) from the best-fit \text{$\Lambda$}CDM model up to $z=1.1$ and extrapolated to higher redshifts. The values of $\sigma$ are redshift-averaged in narrow logarithmic bins. A discrepancy between the data and the model emerges at $z \geq 1$ and it remains significant at higher redshifts. This proves the tension between the \text{$\Lambda$}CDM model and the observations.}
    \label{fig:deltaDM}
\end{figure}
In Figure \ref{fig:deltaDM}, we plot the redshift-averaged $\sigma$ deviations between the cosmographic distance moduli of quasars (yellow bars) and SNe (blue bars) and the best-fit \text{$\Lambda$}CDM model up to $z=1.1$ and extrapolated to higher redshifts. The discrepancy between the \text{$\Lambda$}CDM model and the data is clearly due to a deviation from the Hubble diagram of quasars at $z>1$. This tension increases with redshift, reaching a peak at $z=2-3$ and remaining statistically significant up to the highest redshifts. This is a step forward from the results presented in \citet{lusso19} where the same discrepancy reaches a peak at $z \sim 3$, the redshift of high-quality pointed observations, but it almost vanishes at $z>3$ where the data quality is lower (see their Figure 5). The authors already suggested that observations of quasars at $z \sim 4$ were required to verify this trend. In our sample of quasars new observations at $z>4$ have been included with a tension that is still high at these redshifts, confirming the discrepancy at high redshifts. As a final comment, it is very interesting to notice that quasars are not the only responsible for this discrepancy. In fact, SNe at $z>1$ ($\sim 23$ SNe) already display the same trend and the same evidence for a deviation. As SNe are the most powerful and reliable standard candles in the cosmological analysis, this is a very strong piece of evidence for the existence of a tension between the \text{$\Lambda$}CDM model and the observations. These results, already mentioned in \citet{rl19} and \citet{lusso19}, are now confirmed with greater statistical significance thanks to our new sample of quasars.\\

\begin{figure}
    \resizebox{\hsize}{!}{\includegraphics{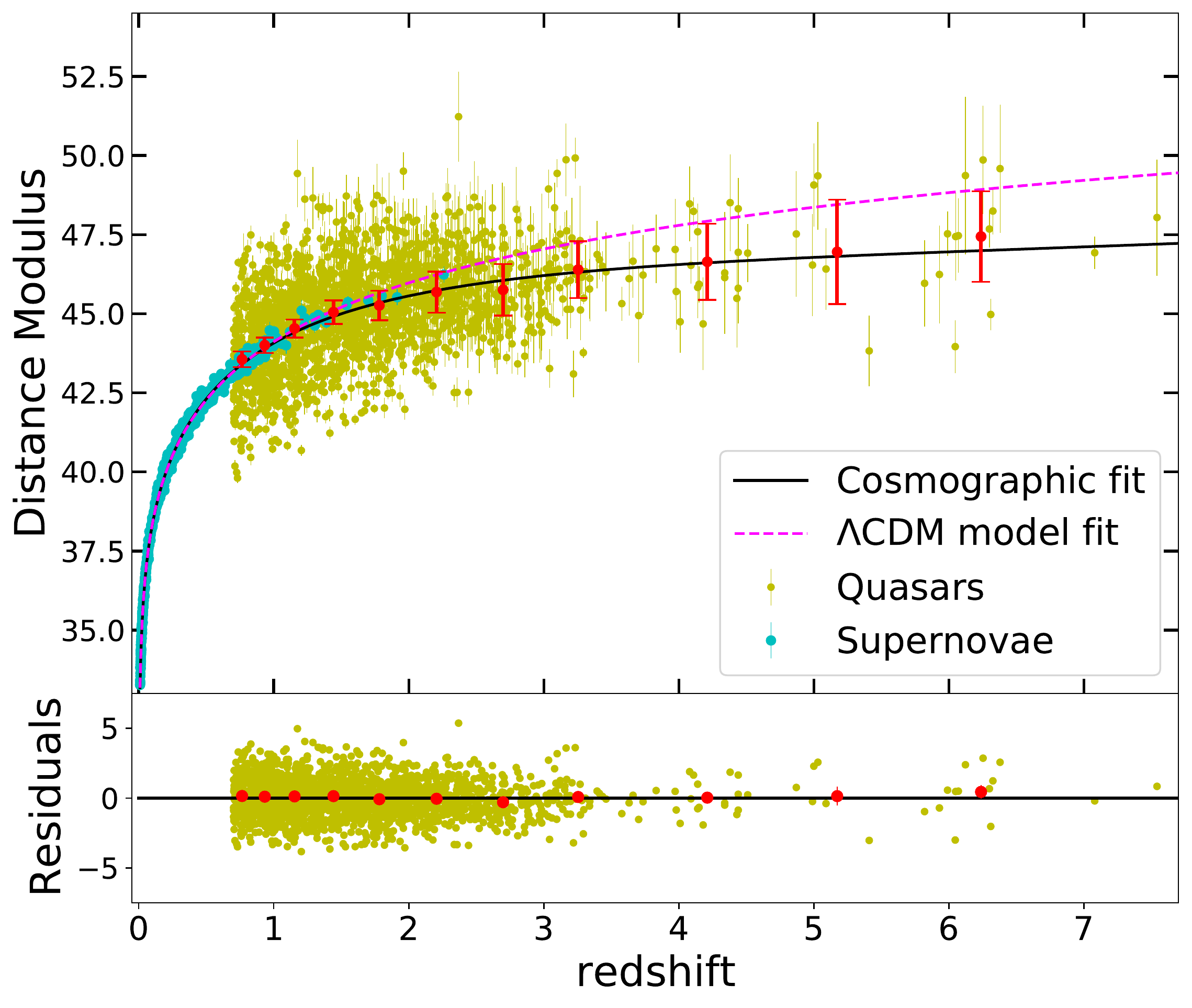}}
    \caption{Top panel: Hubble diagram of SNe (cyan points) and quasars (yellow points). The red points are averages (along with their uncertainties) of the distance moduli of quasars in narrow (logarithmic) redshift intervals and they are shown for visualization purposes only, without any statistical application. Dashed magenta and continuous black lines show respectively the best-fit curves of the flat \text{$\Lambda$}CDM and the logarithmic models over the whole redshift range. Bottom panel: residuals with respect to the cosmographic fit for quasars and averages of the residuals over the same redshift intervals.}
    \label{fig:HD}
\end{figure}

These considerations are reinforced if we analyze the fit of the \text{$\Lambda$}CDM model on the whole redshift range. In Figure \ref{fig:HD} we show the Hubble diagram of quasars (yellow points) and SNe (cyan points) and the best-fit curves of the logarithmic function (continuous black line) and the flat \text{$\Lambda$}CDM model (dashed magenta line) over the whole range of $z$. As the fitting technique requires both the marginalization over all free parameters and the $\sigma$-clipping procedure\footnote{This is a method commonly used to remove possible outliers in the data set. Specifically, it consists of discarding every source which has a discrepancy of more than a specific number of $\sigma$ (in our case $3\sigma$) from the best-fit model. The iteration of this procedure guarantees a better estimation of the free parameters of the model.} on the quasar sample (see \citealt{2020A&A...642A.150L} for more details), we decided to plot the DM values computed through the cosmographic fit. The points corresponding to the \text{$\Lambda$}CDM best-fit model would be different in terms of the number of quasars (due to the $\sigma$-clipping) and DM (due to the marginalization), but this difference is very small and graphically not significant. Indeed, there are only four more quasars, not removed by the $\sigma$-clipping, in the sample produced by the best-fit \text{$\Lambda$}CDM model and the mean value of the difference between the distance moduli ($\text{DM}_{\mathrm{\Lambda CDM}} - \text{DM}_{\text{log}}$) is $\sim 0.3$ with an absolute maximum difference of about 1.3 at $z \sim 3$. This is the reason why, for visual clarity, we have decided to plot both the best-fit curves on the same Hubble diagram (the one from the cosmographic fit), even if this is not formally correct.\\
The main difference with the previous analysis is that the \text{$\Lambda$}CDM model is fitted on the entire redshift range. Comparing with the analysis on the data at $z<1.1$ we notice that: the best-fit values of $\Omega_{M,0}$ are consistent within the error bars ($\Omega_{M,0} = 0.281 \pm 0.013$ for $z<1.1$ and $\Omega_{M,0}=0.296 \pm 0.013$ in the whole range of $z$) and the best-fit curves on the Hubble diagram are indistinguishable with a maximum absolute difference in distance modulus of 0.04.\\
The best-fit curves of the cosmological and cosmographic models shown in Figure \ref{fig:HD} are completely overlapping in the low-redshift region at $z<1$, whilst the tension becomes significant at $z>1.5$, where the prediction of the \text{$\Lambda$}CDM model is much steeper than the one of cosmography.\\
To quantify the discrepancy between these two curves we can compare the best-fit values of $a_{i}$ from the cosmographic fit with the ones predicted from the \text{$\Lambda$}CDM model (see Appendix \ref{appendixcoeff}, relations \ref{coefflcdm}). This comparison is exactly the one between the blue contours and the green star\footnote{As proved in Section \ref{mocksamples}, we are allowed to use the green star instead of the red point as the reference for the prediction of the cosmological model.} already shown in Figure \ref{fig:mock0.28log} for the concordance model ($\Omega_{M,0}=0.28$) and in Figure \ref{fig:mock0.7log} for the case with $\Omega_{M,0}=0.7$. Focusing on the concordance model (Figure \ref{fig:mock0.28log}), as it is the most accepted and successful one, we notice that there is no tension on $a_{2}$, the dominant term at low redshifts. This is a further evidence of the agreement between the \text{$\Lambda$}CDM model and the data up to $z \sim 1$ already shown in the analysis above. Nevertheless, due to the increasing discrepancy on the other parameters and so at higher redshifts, the overall tension between the cosmographic fit and the cosmological model is much greater than 4$\sigma$.  This confirms the results of \citet{rl19} and \citet{lusso19} with evidence of an even increased discrepancy between the standard cosmological model and the data. 


\section{Conclusions}
\label{section5}

We critically analyzed the role of cosmography and its potential issues focusing on the problems that emerge when observational data go beyond $z \sim 1$. We also discussed the possibility to reproduce and fit the data and the standard cosmological model with the final aim of developing a completely model-independent technique to test any cosmological models. Here, we presented a new cosmographic technique based on a polynomial logarithmic function.\\
Firstly, we showed that the new orthogonalization procedure allows us to quantify the discrepancy between the cosmographic fit and any given point in the free parameter spaces without taking into account the possible correlation amongst the free parameters. Moreover, we proved that a fifth-order in the logarithmic polynomial is needed to fit data up to the maximum redshifts of quasars ($z \sim 7.5$), while a sixth-order would not be significant. Then, we studied the behaviour of the logarithmic function in the local range $z \sim 0$ through an expansion centered at $z=0$ and we checked for the possibility to compare this cosmographic function with the prediction of a flat \text{$\Lambda$}CDM model in the same redshift range. The main result was that this expansion is convergent and correctly reconstructs the model. As a consequence, it can be used to fit data and test cosmological models if we limit to $z \leq 1$. After that, we extended this study to the highest redshifts observed ($z \sim 7.5$) through the use of mock samples, as we are not interested in the asymptotic limit where data are not available for now. This analysis showed that an extrapolation to $z>1$ of the logarithmic polynomial with coefficients obtained by the expansion at $z=0$ is robust if we are only interested in a possible tension between the data and the flat \text{$\Lambda$}CDM model. The precision of the cosmographic prediction in the redshift interval $z=1-7.5$ depends on the value of $\Omega_{M,0}$ considered, but it always allows a direct comparison between the cosmographic fit and the prediction of flat \text{$\Lambda$}CDM models with any physical values of $\Omega_{M,0}$.\\
Summarizing this first part of the work, for a cosmographic analysis of the Hubble diagram of quasars and SNe, we need the orthogonal logarithmic approach to obtain a reliable reconstruction of data at $z > 1$ that can be compared with the predictions of cosmological models.\\
Indeed, in the second part of the work, we fitted the Hubble diagram of quasars and SNe with a fifth-order orthogonal logarithmic function and a flat \text{$\Lambda$}CDM model. The main results are:
\begin{itemize}
    \item the Hubble diagram is well reproduced by the \text{$\Lambda$}CDM model up to $z \sim 1$;
    \item a discrepancy emerges at higher redshifts mainly due to quasars, but also due to SNe at $z > 1$, even if with less statistical significance, and it increases with redshift;
    \item the overall tension between cosmographic and cosmological fits is greater than $4\sigma$.
\end{itemize}
In conclusion, we proved that the orthogonalized logarithmic approach we developed is a  powerful and reliable tool to reproduce analytically the Hubble diagram of quasars and SNe at least up to $z \sim 7.5$ and to test the predictions of cosmological models.\\
As we have already pointed out, this work is mainly methodological and it opens the way to analyze the suitability of other cosmographic approaches. In a forthcoming paper, we will apply the method developed to some of Padè polynomials (see \citealt{Benetti} and \citealt{RoccoHigh}). In this case, it would also be possible to include a study on the convergence features of these series at high redshifts, while the convergence of our logarithmic function was already expected due to the logarithmic shape.\\
Moreover, in this paper we only considered a flat \text{$\Lambda$}CDM model as a first application of the orthogonalized cosmographic logarithmic technique. Future works will focus on testing other cosmological models, especially the ones with an evolving dark energy contribution and with a possible interaction between dark energy and dark matter (i.e. Interacting Dark Energy models) or models coming from modified gravity \citep{RoccoReview}\\
A further step  would be to extend the sample used in the cosmological analysis (only quasars and SNe) including other cosmological probes (e.g. CMB, BAO) to study the relationship between early and late time Universe dynamics in the framework of a specific cosmological model.
  
\begin{acknowledgements}
      
\end{acknowledgements}

%

%

\bibliographystyle{aa} 
\bibliography{bibl}

\begin{thebibliography}{41}
\expandafter\ifx\csname natexlab\endcsname\relax\def\natexlab#1{#1}\fi

\bibitem[{{Alestas} {et~al.}(2020{\natexlab{a}}){Alestas}, {Kazantzidis}, \&
  {Perivolaropoulos}}]{2020arXiv201213932A}
{Alestas}, G., {Kazantzidis}, L., \& {Perivolaropoulos}, L. 2020{\natexlab{a}},
  arXiv e-prints, arXiv:2012.13932

\bibitem[{{Alestas} {et~al.}(2020{\natexlab{b}}){Alestas}, {Kazantzidis}, \&
  {Perivolaropoulos}}]{2020PhRvD.101l3516A}
{Alestas}, G., {Kazantzidis}, L., \& {Perivolaropoulos}, L. 2020{\natexlab{b}},
  \prd, 101, 123516

\bibitem[{Aviles {et~al.}(2014)Aviles, Bravetti, Capozziello, \&
  Luongo}]{Aviles}
Aviles, A., Bravetti, A., Capozziello, S., \& Luongo, O. 2014, Phys. Rev., D90,
  043531

\bibitem[{{Ba{\~n}ados} {et~al.}(2018){Ba{\~n}ados}, {Venemans},
  {Mazzucchelli}, {Farina}, {Walter}, {Wang}, {Decarli}, {Stern}, {Fan},
  {Davies}, {Hennawi}, {Simcoe}, {Turner}, {Rix}, {Yang}, {Kelson}, {Rudie}, \&
  {Winters}}]{banados2018}
{Ba{\~n}ados}, E., {Venemans}, B.~P., {Mazzucchelli}, C., {et~al.} 2018, \nat,
  553, 473

\bibitem[{{Banerjee} {et~al.}(2020){Banerjee}, {Colg{\'a}in}, {Sasaki},
  {Sheikh-Jabbari}, \& {Yang}}]{2020arXiv200904109B}
{Banerjee}, A., {Colg{\'a}in}, E.~{\'O}., {Sasaki}, M., {Sheikh-Jabbari},
  M.~M., \& {Yang}, T. 2020, arXiv e-prints, arXiv:2009.04109

\bibitem[{Benetti \& Capozziello(2019)}]{Benetti}
Benetti, M. \& Capozziello, S. 2019, JCAP, 1912, 008

\bibitem[{{Betoule} {et~al.}(2014){Betoule}, {Kessler}, {Guy}, {Mosher},
  {Hardin}, {Biswas}, {Astier}, {El-Hage}, {Konig}, {Kuhlmann}, {Marriner},
  {Pain}, {Regnault}, {Balland}, {Bassett}, {Brown}, {Campbell}, {Carlberg},
  {Cellier-Holzem}, {Cinabro}, {Conley}, {D'Andrea}, {DePoy}, {Doi}, {Ellis},
  {Fabbro}, {Filippenko}, {Foley}, {Frieman}, {Fouchez}, {Galbany}, {Goobar},
  {Gupta}, {Hill}, {Hlozek}, {Hogan}, {Hook}, {Howell}, {Jha}, {Le Guillou},
  {Leloudas}, {Lidman}, {Marshall}, {M{\"o}ller}, {Mour{\~a}o}, {Neveu},
  {Nichol}, {Olmstead}, {Palanque-Delabrouille}, {Perlmutter}, {Prieto},
  {Pritchet}, {Richmond}, {Riess}, {Ruhlmann-Kleider}, {Sako}, {Schahmaneche},
  {Schneider}, {Smith}, {Sollerman}, {Sullivan}, {Walton}, \&
  {Wheeler}}]{betoule2014}
{Betoule}, M., {Kessler}, R., {Guy}, J., {et~al.} 2014, \aap, 568, A22

\bibitem[{{Bullock}(2010)}]{2010arXiv1009.4505B}
{Bullock}, J.~S. 2010, arXiv e-prints, arXiv:1009.4505

\bibitem[{Capozziello {et~al.}(2020{\natexlab{a}})Capozziello, Benetti, \&
  Spallicci}]{Spallicci}
Capozziello, S., Benetti, M., \& Spallicci, A. D. A.~M. 2020{\natexlab{a}},
  Found. Phys., 50, 893

\bibitem[{Capozziello {et~al.}(2018)Capozziello, D'Agostino, \&
  Luongo}]{Chebyshev}
Capozziello, S., D'Agostino, R., \& Luongo, O. 2018, Mon. Not. Roy. Astron.
  Soc., 476, 3924

\bibitem[{Capozziello {et~al.}(2019{\natexlab{a}})Capozziello, D'Agostino, \&
  Luongo}]{RoccoReview}
Capozziello, S., D'Agostino, R., \& Luongo, O. 2019{\natexlab{a}}, Int. J. Mod.
  Phys., D28, 1930016

\bibitem[{Capozziello {et~al.}(2020{\natexlab{b}})Capozziello, D'Agostino, \&
  Luongo}]{RoccoHigh}
Capozziello, S., D'Agostino, R., \& Luongo, O. 2020{\natexlab{b}}, Mon. Not.
  Roy. Astron. Soc., 494, 2576

\bibitem[{Capozziello {et~al.}(2011)Capozziello, Lazkoz, \& Salzano}]{Salzano}
Capozziello, S., Lazkoz, R., \& Salzano, V. 2011, Phys. Rev., D84, 124061

\bibitem[{Capozziello {et~al.}(2019{\natexlab{b}})Capozziello, Ruchika, \&
  Sen}]{Anjan}
Capozziello, S., Ruchika, \& Sen, A.~A. 2019{\natexlab{b}}, Mon. Not. Roy.
  Astron. Soc., 484, 4484

\bibitem[{{Carroll}(2001)}]{2001LRR.....4....1C}
{Carroll}, S.~M. 2001, Living Reviews in Relativity, 4, 1

\bibitem[{{Demianski} {et~al.}(2017){Demianski}, {Piedipalumbo}, {Sawant}, \&
  {Amati}}]{2017A&A...598A.113D}
{Demianski}, M., {Piedipalumbo}, E., {Sawant}, D., \& {Amati}, L. 2017, \aap,
  598, A113

\bibitem[{{Dhawan} {et~al.}(2020){Dhawan}, {Brout}, {Scolnic}, {Goobar},
  {Riess}, \& {Miranda}}]{2020ApJ...894...54D}
{Dhawan}, S., {Brout}, D., {Scolnic}, D., {et~al.} 2020, \apj, 894, 54

\bibitem[{{Escamilla-Rivera} \& {Capozziello}(2019)}]{2019IJMPD..2850154E}
{Escamilla-Rivera}, C. \& {Capozziello}, S. 2019, International Journal of
  Modern Physics D, 28, 1950154

\bibitem[{{Foreman-Mackey} {et~al.}(2013){Foreman-Mackey}, {Hogg}, {Lang}, \&
  {Goodman}}]{2013PASP..125..306F}
{Foreman-Mackey}, D., {Hogg}, D.~W., {Lang}, D., \& {Goodman}, J. 2013, \pasp,
  125, 306

\bibitem[{{Lusso} {et~al.}(2019){Lusso}, {Piedipalumbo}, {Risaliti},
  {Paolillo}, {Bisogni}, {Nardini}, \& {Amati}}]{lusso19}
{Lusso}, E., {Piedipalumbo}, E., {Risaliti}, G., {et~al.} 2019, \aap, 628, L4

\bibitem[{{Lusso} \& {Risaliti}(2016)}]{lr16}
{Lusso}, E. \& {Risaliti}, G. 2016, \apj, 819, 154

\bibitem[{{Lusso} {et~al.}(2020){Lusso}, {Risaliti}, {Nardini}, {Bargiacchi},
  {Benetti}, {Bisogni}, {Capozziello}, {Civano}, {Eggleston}, {Elvis},
  {Fabbiano}, {Gilli}, {Marconi}, {Paolillo}, {Piedipalumbo}, {Salvestrini},
  {Signorini}, \& {Vignali}}]{2020A&A...642A.150L}
{Lusso}, E., {Risaliti}, G., {Nardini}, E., {et~al.} 2020, \aap, 642, A150

\bibitem[{{Peebles} \& {Ratra}(2003)}]{2003RvMP...75..559P}
{Peebles}, P.~J. \& {Ratra}, B. 2003, Reviews of Modern Physics, 75, 559

\bibitem[{{Perlmutter} {et~al.}(1999){Perlmutter}, {Aldering}, {Goldhaber},
  {Knop}, {Nugent}, {Castro}, {Deustua}, {Fabbro}, {Goobar}, {Groom}, {Hook},
  {Kim}, {Kim}, {Lee}, {Nunes}, {Pain}, {Pennypacker}, {Quimby}, {Lidman},
  {Ellis}, {Irwin}, {McMahon}, {Ruiz-Lapuente}, {Walton}, {Schaefer}, {Boyle},
  {Filippenko}, {Matheson}, {Fruchter}, {Panagia}, {Newberg}, {Couch}, \&
  {Project}}]{perlmutter1999}
{Perlmutter}, S., {Aldering}, G., {Goldhaber}, G., {et~al.} 1999, \apj, 517,
  565

\bibitem[{{Planck Collaboration} {et~al.}(2018){Planck Collaboration},
  {Aghanim}, {Akrami}, {Ashdown}, {Aumont}, {Baccigalupi}, {Ballardini},
  {Banday}, {Barreiro}, {Bartolo}, {Basak}, {Battye}, {Benabed}, {Bernard},
  {Bersanelli}, {Bielewicz}, {Bock}, {Bond}, {Borrill}, {Bouchet}, {Boulanger},
  {Bucher}, {Burigana}, {Butler}, {Calabrese}, {Cardoso}, {Carron},
  {Challinor}, {Chiang}, {Chluba}, {Colombo}, {Combet}, {Contreras}, {Crill},
  {Cuttaia}, {de Bernardis}, {de Zotti}, {Delabrouille}, {Delouis}, {Di
  Valentino}, {Diego}, {Dor{\'e}}, {Douspis}, {Ducout}, {Dupac}, {Dusini},
  {Efstathiou}, {Elsner}, {En{\ss}lin}, {Eriksen}, {Fantaye}, {Farhang},
  {Fergusson}, {Fernandez-Cobos}, {Finelli}, {Forastieri}, {Frailis},
  {Franceschi}, {Frolov}, {Galeotta}, {Galli}, {Ganga}, {G{\'e}nova-Santos},
  {Gerbino}, {Ghosh}, {Gonz{\'a}lez-Nuevo}, {G{\'o}rski}, {Gratton},
  {Gruppuso}, {Gudmundsson}, {Hamann}, {Hand ley}, {Herranz}, {Hivon}, {Huang},
  {Jaffe}, {Jones}, {Karakci}, {Keih{\"a}nen}, {Keskitalo}, {Kiiveri}, {Kim},
  {Kisner}, {Knox}, {Krachmalnicoff}, {Kunz}, {Kurki-Suonio}, {Lagache},
  {Lamarre}, {Lasenby}, {Lattanzi}, {Lawrence}, {Le Jeune}, {Lemos},
  {Lesgourgues}, {Levrier}, {Lewis}, {Liguori}, {Lilje}, {Lilley}, {Lindholm},
  {L{\'o}pez-Caniego}, {Lubin}, {Ma}, {Mac{\'\i}as-P{\'e}rez}, {Maggio},
  {Maino}, {Mandolesi}, {Mangilli}, {Marcos-Caballero}, {Maris}, {Martin},
  {Martinelli}, {Mart{\'\i}nez-Gonz{\'a}lez}, {Matarrese}, {Mauri}, {McEwen},
  {Meinhold}, {Melchiorri}, {Mennella}, {Migliaccio}, {Millea}, {Mitra},
  {Miville-Desch{\^e}nes}, {Molinari}, {Montier}, {Morgante}, {Moss}, {Natoli},
  {N{\o}rgaard-Nielsen}, {Pagano}, {Paoletti}, {Partridge}, {Patanchon},
  {Peiris}, {Perrotta}, {Pettorino}, {Piacentini}, {Polastri}, {Polenta},
  {Puget}, {Rachen}, {Reinecke}, {Remazeilles}, {Renzi}, {Rocha}, {Rosset},
  {Roudier}, {Rubi{\~n}o-Mart{\'\i}n}, {Ruiz-Granados}, {Salvati}, {Sandri},
  {Savelainen}, {Scott}, {Shellard}, {Sirignano}, {Sirri}, {Spencer},
  {Sunyaev}, {Suur-Uski}, {Tauber}, {Tavagnacco}, {Tenti}, {Toffolatti},
  {Tomasi}, {Trombetti}, {Valenziano}, {Valiviita}, {Van Tent}, {Vibert},
  {Vielva}, {Villa}, {Vittorio}, {Wand elt}, {Wehus}, {White}, {White},
  {Zacchei}, \& {Zonca}}]{planck2018}
{Planck Collaboration}, {Aghanim}, N., {Akrami}, Y., {et~al.} 2018, arXiv
  e-prints, arXiv:1807.06209

\bibitem[{{Renzi} \& {Silvestri}(2020)}]{2020arXiv201110559R}
{Renzi}, F. \& {Silvestri}, A. 2020, arXiv e-prints, arXiv:2011.10559

\bibitem[{{Rezaei} {et~al.}(2020){Rezaei}, {Pour-Ojaghi}, \&
  {Malekjani}}]{2020ApJ...900...70R}
{Rezaei}, M., {Pour-Ojaghi}, S., \& {Malekjani}, M. 2020, \apj, 900, 70

\bibitem[{{Riess} {et~al.}(2019){Riess}, {Casertano}, {Yuan}, {Macri}, \&
  {Scolnic}}]{riess2019}
{Riess}, A.~G., {Casertano}, S., {Yuan}, W., {Macri}, L.~M., \& {Scolnic}, D.
  2019, \apj, 876, 85

\bibitem[{{Riess} {et~al.}(1998){Riess}, {Filippenko}, {Challis},
  {Clocchiatti}, {Diercks}, {Garnavich}, {Gilliland}, {Hogan}, {Jha},
  {Kirshner}, {Leibundgut}, {Phillips}, {Reiss}, {Schmidt}, {Schommer},
  {Smith}, {Spyromilio}, {Stubbs}, {Suntzeff}, \& {Tonry}}]{riess1998}
{Riess}, A.~G., {Filippenko}, A.~V., {Challis}, P., {et~al.} 1998, \aj, 116,
  1009

\bibitem[{{Riess} {et~al.}(2004){Riess}, {Strolger}, {Tonry}, {Casertano},
  {Ferguson}, {Mobasher}, {Challis}, {Filippenko}, {Jha}, {Li}, {Chornock},
  {Kirshner}, {Leibundgut}, {Dickinson}, {Livio}, {Giavalisco}, {Steidel},
  {Ben{\'\i}tez}, \& {Tsvetanov}}]{2004ApJ...607..665R}
{Riess}, A.~G., {Strolger}, L.-G., {Tonry}, J., {et~al.} 2004, \apj, 607, 665

\bibitem[{{Risaliti} \& {Lusso}(2015)}]{rl15}
{Risaliti}, G. \& {Lusso}, E. 2015, \apj, 815, 33

\bibitem[{{Risaliti} \& {Lusso}(2019)}]{rl19}
{Risaliti}, G. \& {Lusso}, E. 2019, Nature Astronomy, 195

\bibitem[{{Sahni} \& {Starobinsky}(2000)}]{2000IJMPD...9..373S}
{Sahni}, V. \& {Starobinsky}, A. 2000, International Journal of Modern Physics
  D, 9, 373

\bibitem[{{Salvestrini} {et~al.}(2019){Salvestrini}, {Risaliti}, {Bisogni},
  {Lusso}, \& {Vignali}}]{salvestrini2019}
{Salvestrini}, F., {Risaliti}, G., {Bisogni}, S., {Lusso}, E., \& {Vignali}, C.
  2019, \aap, 631, A120

\bibitem[{{Scolnic} {et~al.}(2018){Scolnic}, {Jones}, {Rest}, {Pan},
  {Chornock}, {Foley}, {Huber}, {Kessler}, {Narayan}, {Riess}, {Rodney},
  {Berger}, {Brout}, {Challis}, {Drout}, {Finkbeiner}, {Lunnan}, {Kirshner},
  {Sand ers}, {Schlafly}, {Smartt}, {Stubbs}, {Tonry}, {Wood-Vasey}, {Foley},
  {Hand}, {Johnson}, {Burgett}, {Chambers}, {Draper}, {Hodapp}, {Kaiser},
  {Kudritzki}, {Magnier}, {Metcalfe}, {Bresolin}, {Gall}, {Kotak}, {McCrum}, \&
  {Smith}}]{scolnic2018}
{Scolnic}, D.~M., {Jones}, D.~O., {Rest}, A., {et~al.} 2018, \apj, 859, 101

\bibitem[{{Visser}(2004)}]{visser2004}
{Visser}, M. 2004, Classical and Quantum Gravity, 21, 2603

\bibitem[{{Weinberg}(1972)}]{1972gcpa.book.....W}
{Weinberg}, S. 1972, {Gravitation and Cosmology: Principles and Applications of
  the General Theory of Relativity}

\bibitem[{Weinberg(1989)}]{Weinberg}
Weinberg, S. 1989, Rev. Mod. Phys., 61, 1

\bibitem[{{Wong} {et~al.}(2020){Wong}, {Suyu}, {Chen}, {Rusu}, {Millon},
  {Sluse}, {Bonvin}, {Fassnacht}, {Taubenberger}, {Auger}, {Birrer}, {Chan},
  {Courbin}, {Hilbert}, {Tihhonova}, {Treu}, {Agnello}, {Ding}, {Jee},
  {Komatsu}, {Shajib}, {Sonnenfeld}, {Bland ford}, {Koopmans}, {Marshall}, \&
  {Meylan}}]{2020MNRAS.tmp.1661W}
{Wong}, K.~C., {Suyu}, S.~H., {Chen}, G. C.~F., {et~al.} 2020, \mnras
  [\eprint[arXiv]{1907.04869}]

\bibitem[{{Yang} {et~al.}(2020){Yang}, {Banerjee}, \& {{\'O}
  Colg{\'a}in}}]{2020PhRvD.102l3532Y}
{Yang}, T., {Banerjee}, A., \& {{\'O} Colg{\'a}in}, E. 2020, \prd, 102, 123532

\bibitem[{{Zamora Mun{\~o}z} \& {Escamilla-Rivera}(2020)}]{2020JCAP...12..007Z}
{Zamora Mun{\~o}z}, C. \& {Escamilla-Rivera}, C. 2020, \jcap, 2020, 007

\end{thebibliography}

\appendix

\section{Calculation of the coefficients $k_{ij}$}
\label{appendixk}

The following procedure is the one used to determine the coefficients $k_{ij}$ in order to make all the $a_{i}$ in equation \eqref{Dlog} uncorrelated.\\ 
Firstly, a fit of the data set is performed with a second-order non-orthogonal polynomial $\displaystyle P_{2} = \frac{\text{ln}(10)}{H_{0}}\Bigg[\text{log}(1+z) + a'_{2}\text{log}^{2}(1+z)\Bigg]$ to obtain $a'_{2}$. Then, the same is performed with a third-order non-orthogonal polynomial $\displaystyle P_{3} = \frac{\text{ln}(10)}{H_{0}}\Bigg[\text{log}(1+z) + a''_{2}\text{log}^{2}(1+z) + a''_{3}\text{log}^{3}(1+z)\Bigg]$ to get $a''_{2}$ and $a''_{3}$. The comparison between the orthogonal and the non-orthogonal polynomial expressions leads to the constraint $ a''_{2}=a'_{2}+a''_{3} k_{32}$. This is a step-by-step construction of a polynomial which is formally equivalent to the non-orthogonal one already used in our previous works \citep{rl19,lusso19}, but which has the great advantage of having all the coefficients uncorrelated. Going further with the fourth-order non-orthogonal polynomial $\displaystyle P_{4} = \frac{\text{ln}(10)}{H_{0}}\Bigg[\text{log}(1+z) + a'''_{2}\text{log}^{2}(1+z) + a'''_{3}\text{log}^{3}(1+z) + a'''_{4}\text{log}^{4}(1+z)\Bigg]$ we determine $a'''_{2}$, $a'''_{3}$ and $a'''_{4}$, and with the fifth-order non-orthogonal polynomial $\displaystyle P_{5} = \frac{\text{ln}(10)}{H_{0}}\Bigg[\text{log}(1+z) + a''''_{2}\text{log}^{2}(1+z) + a''''_{3}\text{log}^{3}(1+z) + a''''_{4}\text{log}^{4}(1+z) + a''''_{5}\text{log}^{5}(1+z)\Bigg]$ we obtain $a''''_{2}$, $a''''_{3}$, $a''''_{4}$ and $a''''_{5}$, so we require that
\begin{align}
& a'''_{2}=a'_{2}+a''_{3}k_{32}+a'''_{4}k_{42}\\
& a'''_{3}=a''_{3}+a'''_{4}k_{43}\\
& a''''_{2}=a'_{2}+a''_{3}k_{32}+a'''_{4}k_{42}+a''''_{5}k_{52}\\
& a''''_{3}=a''_{3}+a'''_{4}k_{43}+a''''_{5}k_{53}\\
& a''''_{4}=a'''_{4}+a''''_{5}k_{54}.
\end{align}
The solutions to these six equations are the six unknown $k_{ij}$. The same procedure can be applied to determine the coefficients $k_{ij}$ of the logarithmic orthogonal expansion at any order (not only the fifth one).\\
As this method makes use of the fits on the data set, the fixed coefficients $k_{ij}$ depend on the specific sample considered and the procedure should be repeated for every different sample considered. Nevertheless, as stressed in the main text, this is not a point of particular interest in our case.

\section{Cosmographic logarithmic coefficients in a flat \text{$\Lambda$}CDM model}
\label{appendixcoeff}

The relations between the free parameters $a_{i}$ of the fourth-order cosmographic logarithmic expansion and $\Omega_{M,0}$ in a flat \text{$\Lambda$}CDM model are the following:
\begin{subequations}
\begin{align}
& a_{4}=\text{ln}^{3}(10)\left(-\frac{135}{64} \Omega^{3}_{M,0} + \frac{18}{4} \Omega^{2}_{M,0} - \frac{47}{16} \Omega_{M,0} +\frac{5}{8}\right)\\
& a_{3}=-k_{43}a_{4} + \text{ln}^{2}(10) \left( \frac{9}{8} \Omega^{2}_{M,0} - 2 \Omega_{M,0} + \frac{7}{6} \right)\\
& a_{2}=-k_{32}a_{3}-k_{42}a_{4}+\text{ln}(10) \left( -\frac{3}{4} \Omega_{M,0} + \frac{3}{2} \right) .
\end{align}
\end{subequations}
The extension to the fifth-order is given by
\begin{subequations}
\label{coefflcdm}
\begin{align} 
& a_{5}=\text{ln}^{4}(10) \left( \frac{567}{128} \Omega^{4}_{M,0}-\frac{729}{64}\Omega^{3}_{M,0}+\frac{315}{32}\Omega^{2}_{M,0}-\frac{25}{8}\Omega_{M,0}+\frac{31}{120} \right)\\
& a_{4}=-k_{54}a_{5}+\text{ln}^{3}(10)\left(-\frac{135}{64} \Omega^{3}_{M,0} + \frac{18}{4} \Omega^{2}_{M,0} - \frac{47}{16} \Omega_{M,0} +\frac{5}{8}\right)\\
& a_{3}=-k_{53}a_{5}-k_{43}a_{4} + \text{ln}^{2}(10) \left( \frac{9}{8} \Omega^{2}_{M,0} - 2 \Omega_{M,0} + \frac{7}{6} \right)\\
& a_{2}=-k_{32}a_{3}-k_{42}a_{4}-k_{52}a_{5}+\text{ln}(10) \left( -\frac{3}{4} \Omega_{M,0} + \frac{3}{2} \right) .
\end{align}
\end{subequations}

\end{document}